\documentclass[usenatbib]{mnras}

\usepackage{graphicx}
\usepackage{multirow}
\usepackage{color}
\usepackage{amssymb,amsmath, wasysym,amsfonts}

\newcommand{\ds}{\displaystyle}
\newcommand{\iu}{{i\mkern1mu}}
\newcommand{\kpch}{$h^{-1}\,{\rm kpc}$}
\newcommand{\Mpch}{$h^{-1}\,{\rm Mpc}$}
\newcommand{\kms}{${\rm km~s^{-1}}$}

\newcommand\lsim{\mathrel{\rlap{\lower4pt\hbox{\hskip1pt$\sim$}}
\raise1pt\hbox{$<$}}}
\newcommand\gsim{\mathrel{\rlap{\lower4pt\hbox{\hskip1pt$\sim$}}
\raise1pt\hbox{$>$}}}

\title[Gas rich and gas poor structures]
{Gas rich and gas poor structures through the stream velocity effect} 

\author[Popa et al.]
{Cristina Popa${^1}$\thanks{E-mail:
cpopa@physics.harvard.edu}, Smadar Naoz${^2}$, Federico Marinacci${^3}$,
Mark Vogelsberger${^3}$ \vspace*{0.2cm}\\
  $^1$Physics Department, Harvard University, Cambridge, MA 02138 USA\\
  $^2$Department of Physics and Astronomy, University of California Los Angeles (UCLA), 465 Portola Plaza, Los Angeles, CA 90095\\
  $^3$MIT Kavli Institute for Astrophysics \& Space Research, Cambridge, MA, 02139, USA\\
  }
  
\begin{document}

\pagerange{\pageref{firstpage}--\pageref{lastpage}}
\pubyear{2015}

\maketitle

\label{firstpage}

\begin{abstract}
Using adiabatic high-resolution numerical simulations we quantify the effect of the streaming motion of baryons with respect to dark matter at the time of recombination on structure formation and evolution. Formally a second order effect, the baryonic stream velocity has proven to have significant impact on dark matter halo abundance, as well as on the gas content and morphology of small galaxy clusters. In this work, we study the impact of stream velocity on the formation and gas content of haloes with masses up to $10^9 M_{\astrosun}$, an order of magnitude larger than previous studies. We find that the non-zero stream velocity has a sizable impact on the number density of haloes with masses $\lesssim$ few$\times 10^7 M_{\astrosun}$ up to $z=10$, the final redshift of our simulations. Furthermore, the gas stream velocity induces a suppression of the gas fraction in haloes, which at z=10 is $\sim 10 \%$ for objects with $M\sim10^7M_{\astrosun}$, as well as a flattening of the gas density profiles in the inner regions of haloes. We further identify and study the formation, in the context of a non-zero stream velocity, of moderately long lived gas dominated structures at intermediate redshifts $10 < z < 20$, which Naoz and Narayan have recently proposed as potential progenitors of globular clusters.
\end{abstract}


\section{Introduction}\label{sec:intro}

The current model of structure formation accounts for contributions from a cosmological constant ($\Lambda$), cold dark matter (CDM) and baryons. Prior to recombination ($z \sim 1100$) baryons and photons were tightly coupled and the growth of baryonic over-densities was suppressed. On the other hand, since dark matter is decoupled from the radiation, perturbations that entered the horizon could grow linearly with the scale factor since $z\gsim 3000$. The differences in the history of the evolution of the dark matter and the baryons during this time effectively generated a relative velocity between the two components. This effect was first mentioned in \cite{Sunyaev:1970eu} and its impact on the evolution of the baryonic density fluctuations was studied in a few  theoretical works \citep{Press:1980ApJ, Hu:1995en}. However, since its contribution comes as a second order term in the equations governing the evolution of matter over-densities, it has generally been neglected in both numerical and theoretical studies. 

Recent work by \cite{Tseliakhovich:2010bj} showed that, at the time of recombination, baryons were moving in a coherent manner (on scales of $\lesssim 3$ Mpc comoving) with respect to the dark matter and had an $rms$ velocity of $\sim30$ ${\rm km\,s^{-1}}$. Since, on small scales, the baryons effectively displayed a streaming motion with respect to the dark matter component, their relative velocity is usually referred to as a "stream velocity" in the literature, and we adopt this convention throughout this work. \cite{Tseliakhovich:2010bj} also argued that, since shortly after recombination the temperature of the baryons (and hence the sound speed) drops precipitously, the streaming motion of the baryons becomes supersonic and it can potentially have a dramatic impact on structure formation. 
Subsequent studies, both theoretical and numerical, showed that the large stream velocity of baryons has profound implications on the 
total number density of haloes \citep{Asaba:2015hwa, Tanaka:2013bxg, Tanaka:2013boa, Bovy:2012af, O'Leary:2012rs, Naoz:2011if, Fialkov:2011iw, Tseliakhovich:2010bj, Tseliakhovich:2010yw, Maio:2010qi} 
the size of the haloes that are able to retain gas at a given redshift \citep{Naoz:2012fr}, 
the overall gas fraction in haloes \citep{Asaba:2015hwa, Richardson:2013uqa, Maio:2010qi, O'Leary:2012rs, Naoz:2012fr, Greif:2011iv, Fialkov:2011iw, Naoz:2011if, Tseliakhovich:2010yw, Dalal:2010yt}, 
the gas density and temperature profiles \citep{Richardson:2013uqa, O'Leary:2012rs, Fialkov:2011iw, Greif:2011iv, Liu:2010xb}, 
the halo mass threshold in which star formation occurs \citep{Bovy:2012af,O'Leary:2012rs, Fialkov:2011iw, Greif:2011iv, Maio:2010qi, Liu:2010xb}, as well as 
black hole evolution \citep{Latif:2013oga, Tanaka:2013boa, Tanaka:2013bxg} and 
magnetic fields \citep{Naoz:2013wla}. A thorough review of the implications of the stream velocity on structure formation is presented in \cite{Fialkov:2014rba}.

A consensus of previous studies accounting for the relative motion of the baryons with respect to the dark matter is that a non-zero stream velocity correlates with a decrease in the number density of small haloes with masses $M \lesssim 10^7 M_{\astrosun}$, as well as with a decrease in the gas fraction contained in these objects. However, since the baryonic fraction is roughly constant when computed over large portions of space, a natural question which arises is, therefore, what happened to the gas which did not end up inside dark matter haloes. \cite{Naoz:2014bqa} suggested that, a non-zero stream velocity correlates with a physical offset between the centers of the dark matter and the baryon over-densities. Therefore, the baryon and dark matter over-densities are expected to collapse at slightly different locations, potentially forming distinct dark matter dominated and baryon dominated objects. Furthermore, \cite{Naoz:2014bqa} suggest that, while in many cases the collapsed baryon over-densities will either be absorbed into a nearby dark matter halo or evaporate, some objects with gas masses $10^5M_{\astrosun} < M_{gas} < $ few $\times 10^6M_{\astrosun}$ may survive outside dark matter haloes and become the progenitors of globular clusters observed today. Their dark matter counterpart would, in turn, be gas-poor and it might become a dark satellite or an ultra faint galaxy. 

In this paper, we perform a suite of high-resolution numerical simulations to study the effect of the stream velocity on structure formation, focusing on the evolution of baryonic structures. We consider three distinct scenarios in which the relative velocity of the baryons with respect to the dark matter component at the time of recombination is $0$ (the baseline commonly used in the literature), equal to the value of its root mean squared $\sigma_{v_{bc}}$ (moderate velocity) and equal to $2\times\sigma_{v_{bc}}$ (high velocity) and evolve them adiabatically from redshift 200 to redshift 10. We aim to systematically quantify the impact of the stream velocity on objects with masses up to $10^9M_{\astrosun}$ as well as investigate the probability of forming long lived, gas dominated objects, as predicted by \cite{Naoz:2014bqa}.

The rest of the paper is structured as follows. In Section \ref{sec:init} we explain the theoretical foundation of the setup of the initial conditions for our simulation, followed by a description of the parameters of the simulations performed in Section \ref{sec:runParams} and the structure definitions used in interpreting our results in Section \ref{sec:halodef}. We discuss the impact of the stream velocity on the number density of dark matter haloes recovered in our simulations in Section \ref{sec:DMp} and on their gas content in Section \ref{sec:DMpGASs}. Finally, we present the properties of the gas rich objects identified in our simulations in Section \ref{sec:GASp} and conclude with a discussion in Section \ref{sec:discussion}.

\section{Simulation Setup}

\subsection{Initial Conditions}\label{sec:init}

The initial conditions used in numerical simulations can have a significant impact on the formation of the first structures, as well as on the properties of haloes at formation (such as their abundance, gas fraction and, implicitly, on the onset of reionization e.g. \citealt{Naoz:2006tr, Naoz:2006ye, Yoshida:2003sy, Naoz:2010hg}). Following \cite{Naoz:2010hg} we generate two separate transfer functions for dark matter and baryons, ensuring that the streaming velocities are realized in a self-consistent way with the transfer functions. The evolution of the dark matter and baryon over-densities  can be written as
\begin{eqnarray} \label{eq:compactFinaldm}
 \ddot{\delta}_{c} + 2 H \dot{\delta}_c -  \frac{2\iu}{a}\mathbf{v_{bc}}\cdot\mathbf{k}\dot{\delta}_c  &= & \frac{3}{2} H_0^2 \Omega_m(f_b \delta_b + f_c \delta_c)  \nonumber \\
&+& \left(\frac{\mathbf{v_{bc}} \cdot \mathbf{k}}{a}\right)^2 \delta_c   \\
\ddot{\delta}_{b} + 2 H  \dot{\delta}_{b}& =& \frac{3}{2} H_0^2 \Omega_m(f_b \delta_b + f_c \delta_c) \nonumber \\
&-& \frac{ k^2 } {a^2} \frac{k_B \bar{T}}{\mu}(\delta_b+\delta_T) \label{eq:compactFinalb}
\end{eqnarray}
Note that the temperature term in Equation (\ref{eq:compactFinalb}) is a first order correction to the linear evolution, which accounts for scale-dependent temperature fluctuations \citep{Naoz:2005pd}. The terms proportional to $v_{\rm bc}$ are second order terms that account for the streaming velocity  \citep{Tseliakhovich:2010bj, Tseliakhovich:2010yw}. Furthermore, as described in \cite{Naoz:2005pd}, using the equation of state and the first law of thermodynamics, one can relate the evolution of the the temperature fluctuations to the density fluctuations by 
\begin{align}
& \dot{\delta}_T =  \frac{2}{3}\dot{\delta}_b + \frac{x_e(t)}{t_\gamma} \left[\delta_\gamma\left(\frac{\bar{T}_\gamma}{\bar{T}}-1\right) + \frac{\bar{T}_\gamma}{\bar{T}}( \delta_{T_\gamma} -\delta_T)\right] a^{-4} \ .
\end{align}
In the above equations $\Omega_m$ is the present day matter density, $H_0$ is the present day value of the Hubble constant $H$, $f_b$ and $f_c$ are the cosmic baryon and dark matter fractions, respectively, $\delta_b$, $\delta_c$, $\delta_\gamma$, $\delta_T$ and $\delta_{T_\gamma}$ are the fluctuations of the dark matter density,  baryon density,  photon density,  mean baryon temperature $\bar{T}$ and  mean photon temperature $T_\gamma$, $x_e(t)$ is the electron fraction out of the total number density of gas particles, $k$ is the comoving wavenumber, $a$ is the scale factor, $\mu$ is the mean molecular weight, $ t_\gamma^{-1}  = 8.55\times10^{-13}$ yr$^{-1}$ and $v_{bc}$ is the average relative velocity between the baryons and the dark matter in a given region of space. Note that in order to solve the above equations for the evolution of the density and temperature fluctuations one needs to assume a value for the stream velocity as well as an angle with respect to the $\mathbf{k}$ vector. Therefore, any such solutions are only valid in a region of space of the order of the coherence scale of the relative velocity of the baryons and the dark matter, roughly $\lesssim 3$ Mpc comoving \citep{Tseliakhovich:2010bj, Tseliakhovich:2010yw}. 

\begin{table*}
\centering
\begin{tabular}{| l | r | l | l | l | l |}
\hline
Box size 		& $v_{bc}$ 		& $\sigma_{v_{bc}}$	& No of particles	& $m_{DM}$					& $m_{gas}$	 \\
			& (at $z=200$) 		&  				&				& 							&		 \\\hline
\multirow{3}{*}{700 \kpch}	& 0 \kms		& $0\times\sigma_{v_{bc}}$	& \multirow{3}{*}{$2\times 512^3$}	& \multirow{3}{*}{$230 M_{\astrosun}$}			& \multirow{3}{*}{$46 M_{\astrosun}$	}\\
						& 5.8 \kms	& $1\times\sigma_{v_{bc}}$	&							&										& 	\\
						& 11.8 \kms	& $2\times\sigma_{v_{bc}}$	&							&										&	\\\hline
\multirow{2}{*}{1.12 \Mpch}	& 0 \kms		& $0\times\sigma_{v_{bc}}$	& \multirow{2}{*}{$2\times 512^3$}	& \multirow{2}{*}{$950 M_{\astrosun}$}	& \multirow{2}{*}{$186 M_{\astrosun}$} \\
						& 11.8 \kms	& $2\times\sigma_{v_{bc}}$	&							&										&	\\\hline
\multirow{2}{*}{1.4 \Mpch}	& 0 \kms		& $0\times\sigma_{v_{bc}}$	& \multirow{2}{*}{$2\times 512^3$}	& \multirow{2}{*}{$1.9\times10^3 M_{\astrosun}$}	& \multirow{2}{*}{$360 M_{\astrosun}$} \\
						& 11.8 \kms	& $2\times\sigma_{v_{bc}}$	&							&										&	\\\hline
\multirow{2}{*}{2.8 \Mpch}	& 0 \kms		& $0\times\sigma_{v_{bc}}$	& \multirow{2}{*}{$2\times 768^3$}	& \multirow{2}{*}{$4.4\times10^3 M_{\astrosun}$}	& \multirow{2}{*}{$843 M_{\astrosun}$}	\\
						& 11.8 \kms	& $2\times\sigma_{v_{bc}}$	&							&	&	\\\hline
\end{tabular}
\caption{Details of the different physics scenarios and resolutions of the simulations run. The zero-stream velocity and the high stream velocity scenarios are explored at all resolution levels, while the moderate stream velocity case is only explored in the $700$ \kpch, highest resolution run.}\label{table:runs}
\end{table*}

We use a modified version of the  {\sc cmbfast} code \citep{Seljak:1996is} to generate the initial conditions at the time of recombination. The code solves the coupled second order equation (\ref{eq:compactFinaldm}) and (\ref{eq:compactFinalb}),  accounting for the stream velocity $\mathbf{v}_{bc}$ and the inhomogeneous speed of sound (see \citealt{Naoz:2005pd}, for the implementation of this term) and creates separate transfer functions  for the baryonic and the dark matter components at the initial redshift 200. Furthermore, assuming that the relative velocity of the baryons and the dark matter can be approximated with a stream velocity for all of our simulations, in setting up our initial conditions, we impart on the baryons a constant velocity boost in the $x$ direction equivalent to the $v_{bc}$ used to generate the transfer function for the initial density fluctuations.

\subsection{Run parameters}\label{sec:runParams}

We run a suite of {\sc arepo} \citep{Springel:2009aa} simulations starting at redshift 200 and evolve them adiabatically to redshift 10. The details of the simulation runs together with their dark matter and gas mass resolution are shown in Table \ref{table:runs}. 

We are investigating zero-stream velocity ($\mathbf{v}_{bc} = 0$) and high-stream velocity ($\mathbf{v}_{bc} = 2\times\sigma_{v_{bc}}$) scenarios at four different resolution levels, as well as an intermediate-stream velocity case ($\mathbf{v}_{bc} = \sigma_{v_{bc}}$) in our highest resolution runs. 
Throughout the paper, we express the value of the stream velocity in units of its $rms$ ($\sigma_{v_{bc}}$), in order to avoid any confusion regarding its value. 
At recombination $\sigma_{v_{bc}} \sim 30$ km/s, as discussed in \cite{Tseliakhovich:2010bj}, and at the beginning of our numerical simulations, at redshift 200, $\sigma_{v_{bc}}\sim 5.8$ km/s. 
For our three highest resolution simulations, the box sizes used ($700$ \kpch, $1.12$ \Mpch\ and $1.4$ \Mpch) lie well below the coherence length of the relative velocity of baryons and dark matter and the relative motion can be approximated with a stream velocity. However, the smaller the size of the simulation box, the less massive the haloes that will form at the final redshift. Hence, in order to access a larger halo mass range we chose to perform a $2.8$ \Mpch\ simulation as well. The size of this box lies at the very edge of the coherence length for the relative velocity and, as we discuss later in the paper, this leads to an overestimation of the $v_{bc}$ effect particularly for small haloes. Yet, since properties of intermediate to large mass haloes in this simulation converge to those recovered in our finer resolution runs, we do find the predictions of the $2.8$ \Mpch\ simulations for the high halo mass region to be reliable. 

For all simulations we are considering a $\Lambda$CDM cosmology with $\Omega_{\Lambda} = 0.73, \Omega_m = 0.27, \Omega_B = 0.044$ and $h = 0.71$. Since the box sizes of our simulations are limited by the coherence length of the stream velocity of baryons, we artificially increase the gravitational clustering in the simulation by setting $\sigma_8 = 1.7$, in order to compensate for the missing large-scale power. We choose these parameters in order to be able to resolve with at least 300 gas particles haloes with $\ds M_{gas} > 1.4 \times 10^{4} M_{\astrosun}$ for the 700 \kpch\ runs, $M_{gas} > 5.6\times10^4 M_{\astrosun}$ for the $1.12$ \Mpch\ runs, $M_{gas} > 1.1\times10^5 M_{\astrosun}$ for the $1.4$ \Mpch\ runs and $M_{gas} > 2.6\times10^5 M_{\astrosun}$ for the $2.8$ \Mpch\ runs. This will allow us to accurately measure the gas fraction in gas-rich regions \citep{Naoz:2009zu} and potentially identify the gas-dominated structures predicted by \cite{Naoz:2014bqa}, which might form due to the presence of a non-zero stream velocity. 
 
 \begin{figure*}
\centering
\begin{minipage}[h]{.47\textwidth}
\centering
\includegraphics[scale=.45]{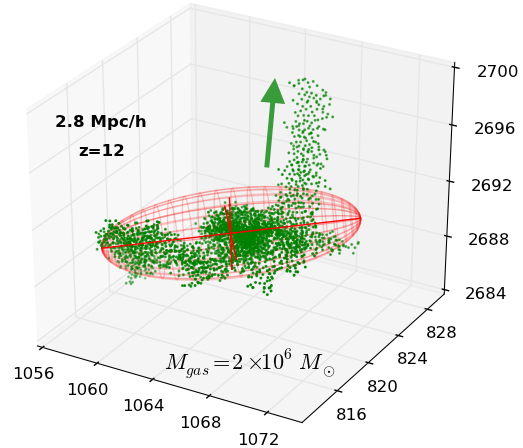}
\end{minipage}\qquad
\begin{minipage}[h]{.47\textwidth}
\centering
\includegraphics[scale=.45]{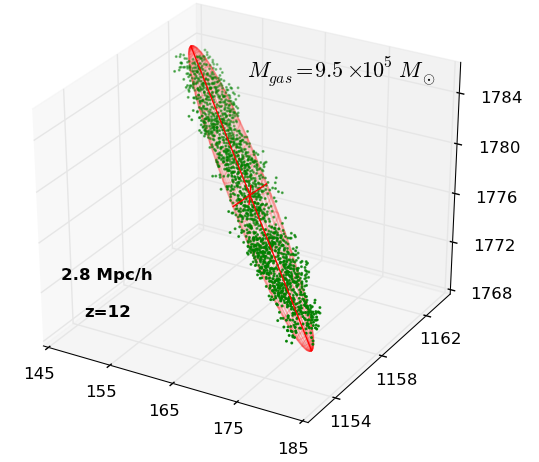}
\end{minipage}
\qquad\caption{The gas content of two Gas-Primary objects identified in the high stream 
velocity, 2.8 \Mpch\ simulation at redshift 20. Many of the Gas-Primary structures 
identified have a  filamentary shape and, in studying their properties, 
spherical symmetry is not a valid assumption anymore. The red surfaces represent 
 tightly fitting ellipsoids which we use in order to define the boundaries of 
these gas abundant structures, so as to exclude sparsely populated regions of the 
FOF groups, such as the upward streaming tail of the object in the left panel. Because 
we fit these surfaces to mold closely to the gas dominated region, the gas fraction 
inside the ellipsoids will be biased high. The green arrow in the left panel represents the direction of the flow of particles into a nearby dark matter halo.}
\label{fig:3dEllipsoids}
\end{figure*}

For all the runs, glass-like adiabatic initial conditions were generated using the Zel'dovich approximation, with the transfer functions computed as described in Section \ref{sec:init}. For the baryons, we used a glass file with positions shifted by a random vector, thus removing all artificial coupling between DM and gas \citep{Yoshida:2003sy}. We set the initial temperature of the gas to be $422\, {\rm K}$, as derived from linear theory, and we use throughout a softening length of $30$ $h^{-1}\,{\rm pc}$ for both the gas and dark matter. 

\subsection{Structure definition}\label{sec:halodef}

The halo definitions used in our analysis are motivated by the scope of this work. On the one hand, we are interested in understanding the impact of a non-zero stream velocity on the growth of dark matter over-densities and, on the other hand, we want to study the evolution of gas over-densities independent of the dark matter component. The most commonly used technique to identify structure in N-body simulations is the  friends-of-friends (FOF) halo finding algorithm. Given a collection of particles, it assigns any two particles with a separation smaller than a specified linking length $b$ as belonging to the same group. By running a FOF halo finding algorithm on the dark matter component of our simulations we identify {\emph{Dark Matter-Primary}} objects and by running the same algorithm using solely the information regarding the baryonic component, we identify {\emph {Gas-Primary}} objects. Finally, we also use the FOF finding algorithm to identify {\emph {Dark Matter-Primary, Gas-Secondary}} structures by constructing dark matter objects to which we attach gas cells in a secondary linking stage \citep{Dolag:2008ar}. Note that the names DM-Primary, Gas-Primary and DM-Primary, Gas-Secondary refer to the particle type used as primary linking particle or attached in a secondary stage by the FOF halo finding algorithm, and we expect all identified objects to contain both dark matter and gas. Throughout our analysis, we use a linking length of 0.03 times the mean particle separation in selecting FOF groups. 

We use the information about the DM-Primary structures identified to understand the impact of a non-zero stream velocity on the formation and evolution of dark matter haloes, regardless of their gas content. There is a direct correspondence between DM-Primary structures and DM-Primary, Gas-Secondary structures and we use the latter to study the impact of the stream velocity on the gas content of the dark matter haloes. For the rest of the paper, we assume spherical symmetry for both the DM-Primary structures, as well as the DM-Primary, Gas-Secondary structures and we identify the virial radius of these haloes as the radius of a sphere centered on the center of mass of the objects, enclosing a density 200 times the critical matter density of the universe. 

Finally, we identify Gas-Primary structures by running the FOF halo finding algorithm using gas cells as primary linking particles. These objects are centered on the gas over-densities in our simulations and, unlike the haloes identified based on the dark matter particle information, they can often have filamentary shapes, as shown in Figure \ref{fig:3dEllipsoids}. Therefore, in studying their properties we relax the assumption of spherical symmetry to ellipsoidal symmetry. Using all of the gas cells in the FOF group, we identify an ellipsoidal surface enclosing all of the gas cells in the FOF group. However, since these structures can be potentially very sparse, we proceed to identifying a "tightly fitting ellipsoid" using the following iterative procedure. 
By progressively shrinking the lengths of the axes of the original ellipsoid enclosing all of the gas cells in the Gas-Primary object in decrements of 0.5\% of their initial value, we stop when the ratio of the lengths of the axes of the small ellipsoid to those of the original ellipsoid is greater than the ratio of the number of gas cells enclosed in the small ellipsoid to the number of gas cells in the original ellipsoid. In this way, we make sure not to artificially remove gas cells from high density regions of the group. In addition, we require to eliminate not more than 20\% of the total number of gas cells present in the original FOF group.  In our analysis below, it is the matter content of this tightly fitting ellipsoid that we study when analyzing Gas-Primary objects. We note that even though the Gas-Primary objects are defined by running the FOF halo finding algorithm only using the information related to the gas cells, the ellipsoidal surfaces  will generally contain both dark matter and baryons. Also, many of the Gas-Primary structures are, in fact, simply the gas components of the DM-Primary, Gas-Secondary haloes. However, since in one case we analyze the matter content of the object using tightly fitting ellipsoids centered on the center of mass of the gas and in the other case we use a sphere of viral radius centered on the center of mass of the dark matter + gas object, the gas fraction found for the former will naturally be biased higher than the latter. 

\begin{figure*}
\centering
\begin{minipage}[b]{.47\textwidth}
\centering
\includegraphics[scale=.4]{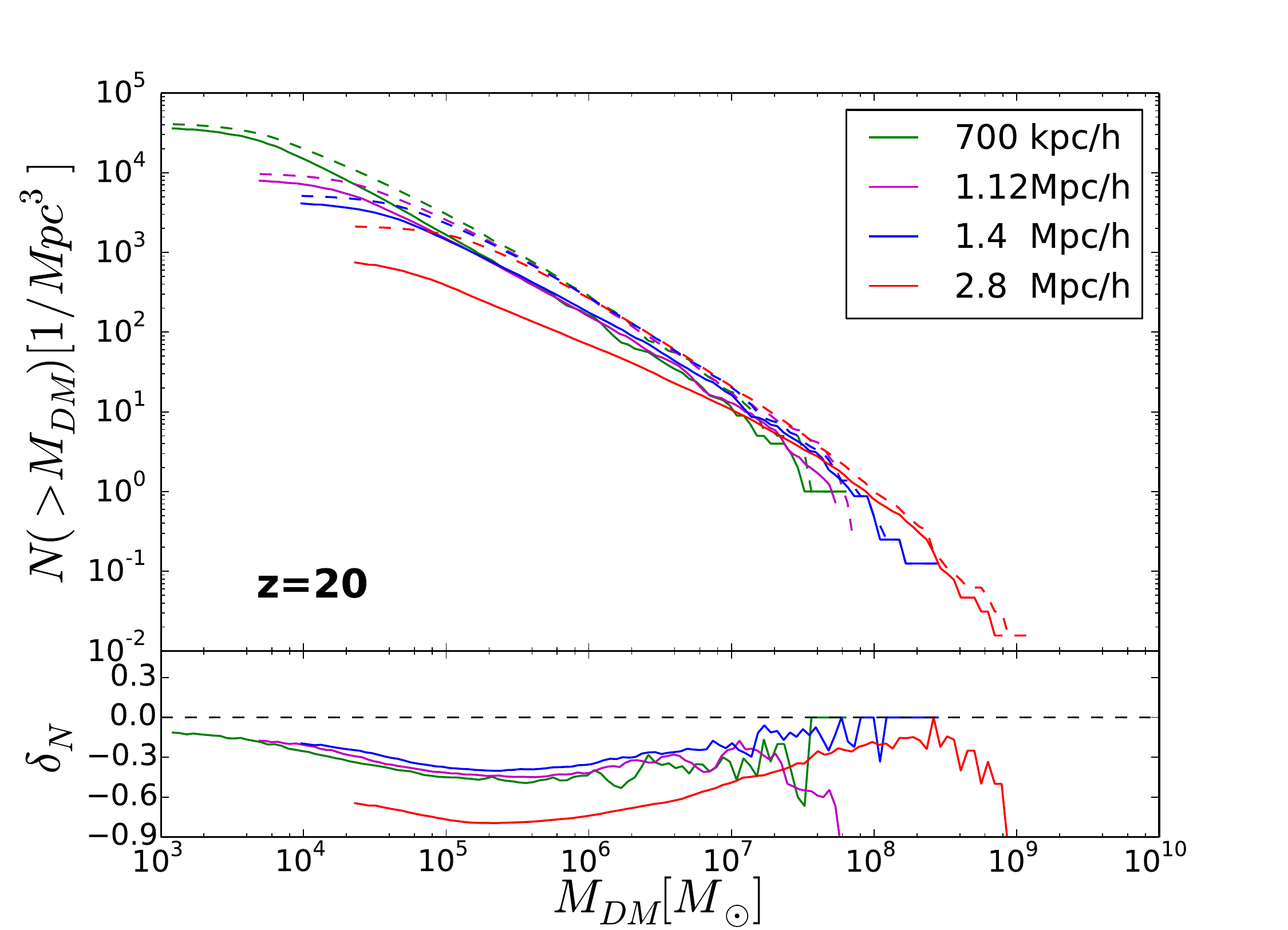}
\end{minipage}\qquad
\begin{minipage}[b]{.47\textwidth}
\centering
\includegraphics[scale=.4]{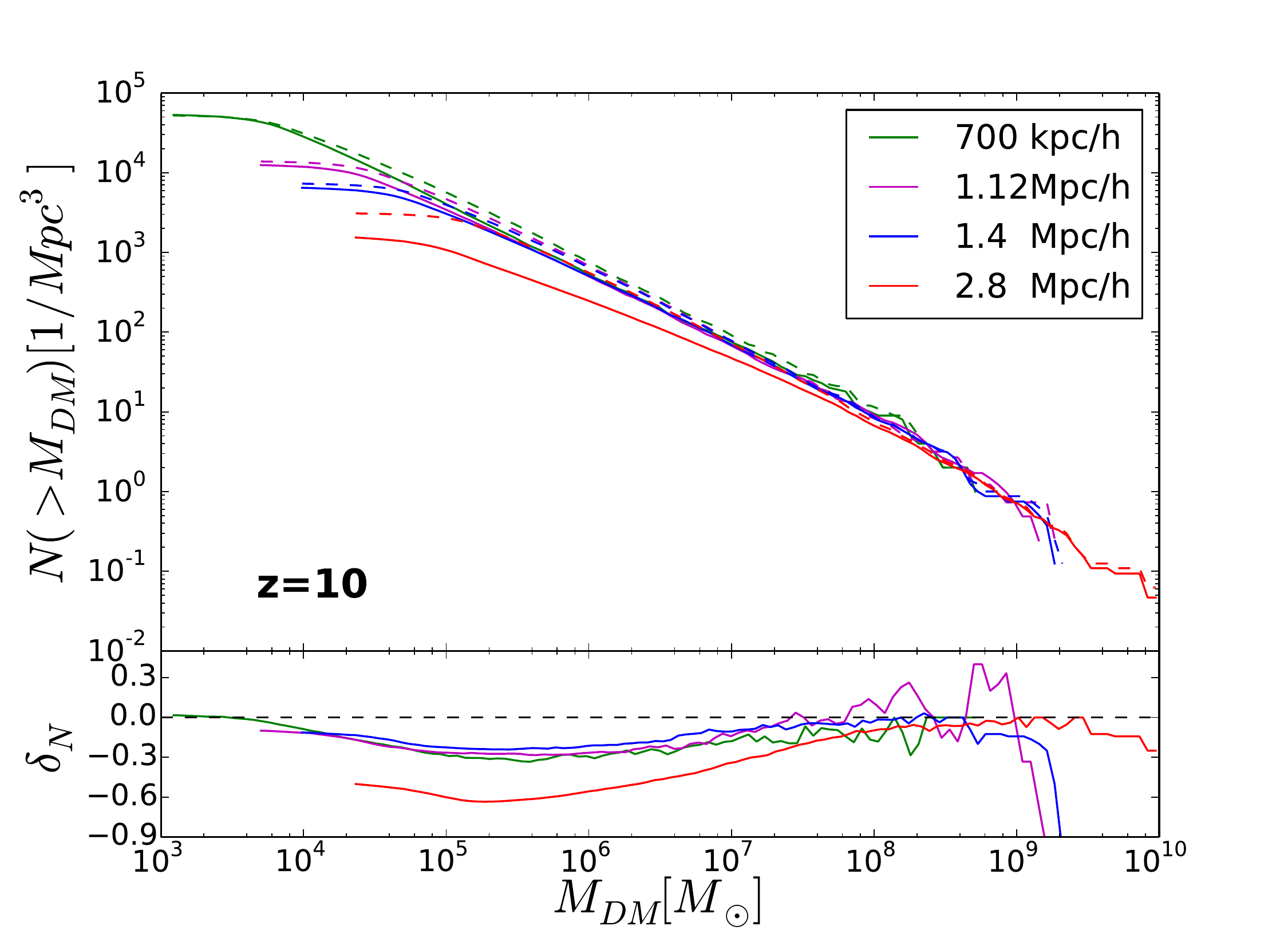}
\end{minipage}\qquad
\caption{Top panels: Cumulative halo mass functions of the DM-Primary objects for the $v_{bc}= 0$ (dashed line) and $v_{bc} = 2\times\sigma_{v_{bc}}$ (solid line) simulations at redshifts 20 (left) and 10 (right) for the different resolutions considered. Bottom panels: Fractional difference in the number density of haloes $\ds\left(\delta_N = (N_{2\sigma} - N_0) / N_0 \right)$ as a function of halo mass. }
\label{fig:haloMassFnc-DM}
\end{figure*}

For all the different structure definitions mentioned above -- DM-Primary, Gas-Primary and DM-Primary, Gas-Secondary -- we only consider FOF groups that contain more than 300 dark matter particles and/or 300 gas particles. This, criterion, on the one hand excludes objects that are not very massive, and, on the other hand, ensures that the gas and dark matter mass computations have an accuracy of the order of $10-20\%$ \citep{Naoz:2009zu}.

\section{Results}

We investigate the impact of the streaming motion of the baryons with respect to dark matter on structure formation and evolution at intermediate redshifts between $z=20$ and $z=10$. We are interested in quantifying the effect that the stream velocity has on the number density of haloes of different masses, as well as its redshift dependence. We explore these issues in Section \ref{sec:DMp}. We continue our study of the impact of the stream velocity on dark matter haloes by studying the gas fraction as a function of halo mass in Section \ref{sec:DMpGASs}. Furthermore, since the numerical simulations of different box sizes effectively access different halo mass ranges, we use this analysis to study the convergence of the 4 different resolution levels. Finally, in Section \ref{sec:GASp} we investigate the potential for long lived gas rich objects to form and evolve outside dark matter haloes, as predicted by \cite{Naoz:2014bqa}.

\subsection{Suppression of halo mass function}\label{sec:DMp}

Understanding the halo abundance in the early Universe is of crucial importance to linking observations of galaxy cluster parameters to a specific set of cosmological parameters. Previous semi-analytical studies  \citep{Tseliakhovich:2010bj, Fialkov:2011iw, Bovy:2012af, Tseliakhovich:2010yw, Tanaka:2013bxg, Tanaka:2013boa} as well as numerical simulations \citep{Maio:2010qi, O'Leary:2012rs, Naoz:2011if} have found that a non-zero stream velocity correlates with a decrease of the halo mass function, particularly for small haloes ($M \lesssim 10^7 M_{\astrosun}$) at high redshifts ($z>20$). \cite{Tseliakhovich:2010bj}, using an argument based on the Press-Schecter formalism, showed that the number density of haloes with mass $M \sim 10^6M_{\astrosun}$ is suppressed by 60\% at redshift $40$ in regions with $v_{bc}\sim\sigma_{v_{bc}}$, compared to regions with no stream velocity. These results are consistent with numerical simulations. \cite{Naoz:2011if} found that $v_{bc} = 3.4\sigma_{v_{bc}}$ simulations exhibit a 50\% suppression and $v_{bc} = \sigma_{v_{bc}}$ simulations exhibit a 15\% suppression in the number density of haloes with $M \simeq 10^5 M_{\astrosun}$, at $z=19$.

\begin{figure*}
\centering
\begin{minipage}[b]{.4\textwidth}
\centering
\includegraphics[scale=.4]{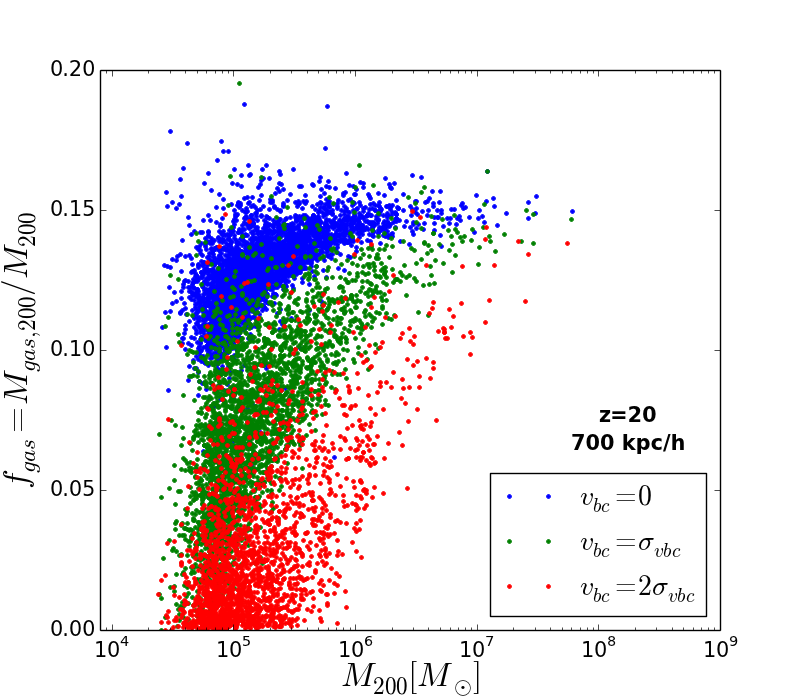}
\end{minipage}\qquad
\begin{minipage}[b]{.4\textwidth}
\centering
\includegraphics[scale=.4]{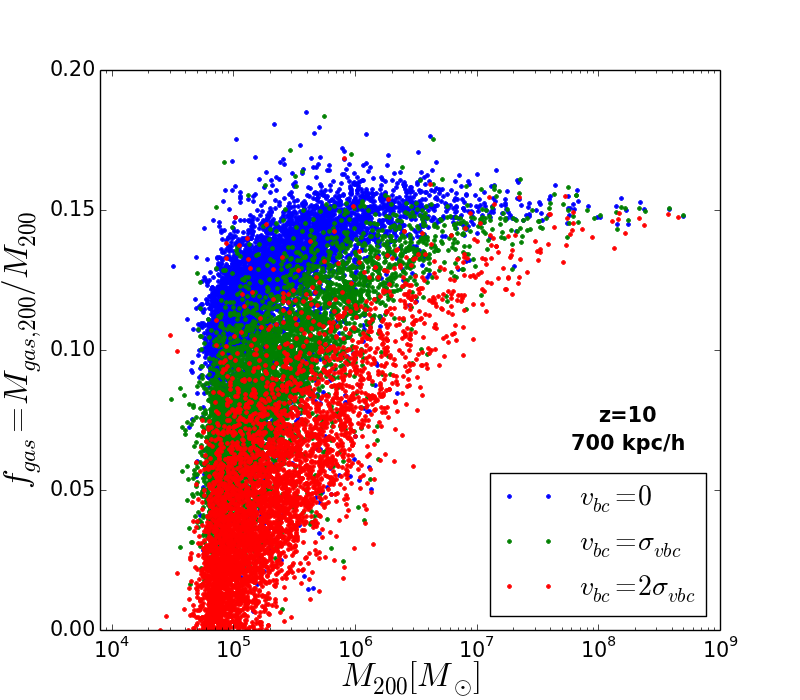}
\end{minipage}\qquad
\begin{minipage}[b]{.4\textwidth}
\centering
\includegraphics[scale=.4]{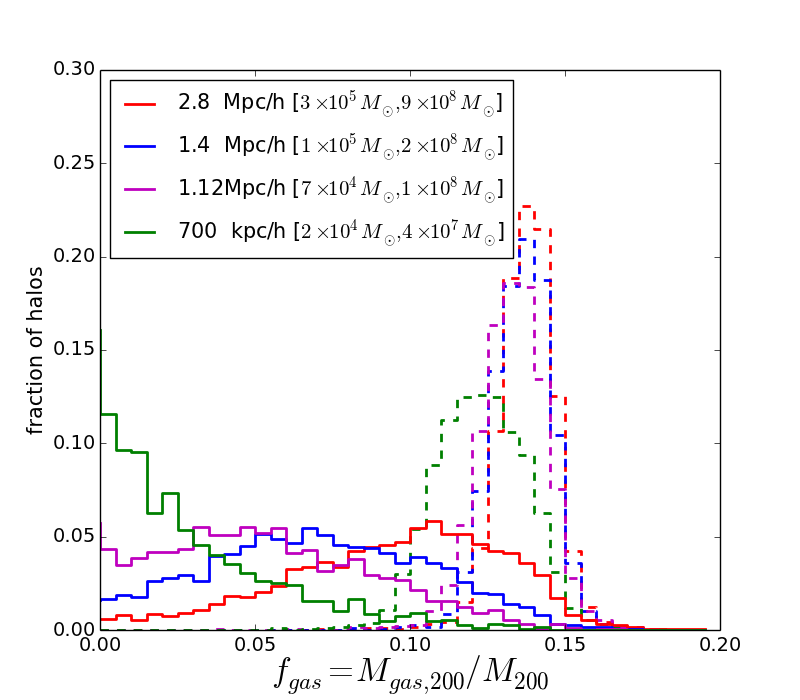}
\end{minipage}\qquad
\begin{minipage}[b]{.36\textwidth}
\centering
\includegraphics[scale=.4]{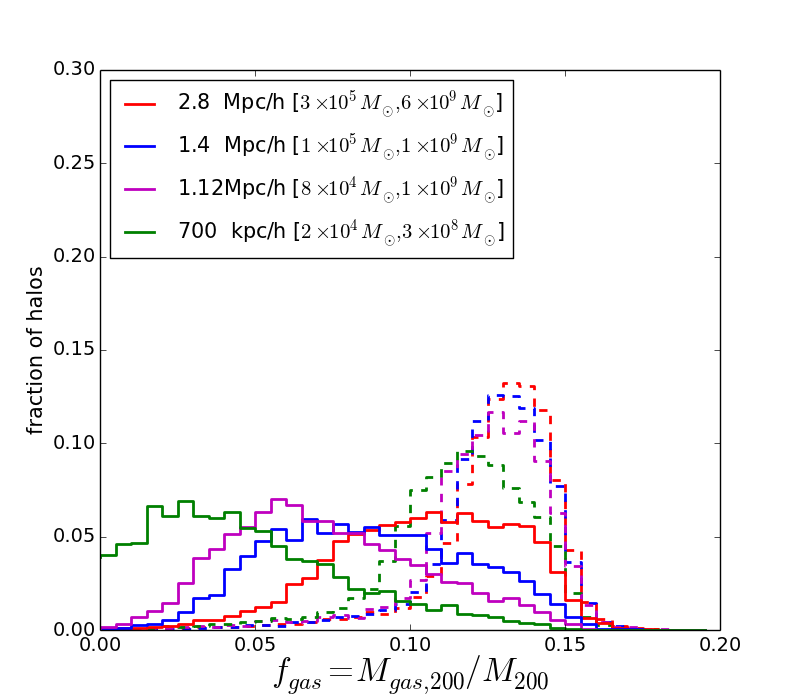}
\end{minipage}
\qquad\caption{Top: Scatter plots of the gas fraction as a function of total mass for the DM-primary, Gas-secondary objects identified in the $700$ \kpch\ simulation. See Figure \ref{fig:rhofgasDMHalos1} for similar scatter plots from the $1.12$ \Mpch\, $1.4$ \Mpch\ and $2.8$ \Mpch\ runs. Bottom: Histograms normalized to unity of the gas fraction in DM-primary, Gas-secondary objects identified in the $v_{bc} = 0$ (dashed lines) and the $v_{bc} = 2\times\sigma_{v_{bc}}$ (solid lines) runs, at different resolutions. The different resolution runs effectively access different total mass ranges for the DM-primary, Gas-secondary objects, as indicated in the legend. } 
\label{fig:rhofgasDMHalos}
\end{figure*}

Our simulations are consistent with these results and, furthermore, show that the suppression of the halo mass function persists up to redshift 10. This observation holds true for both DM-primary objects, as well as  DM-primary, Gas-secondary objects. Figure \ref{fig:haloMassFnc-DM} shows the halo mass function computed at different resolutions for the simulations with $v_{bc} = 2\times\sigma_{v_{bc}}$ (solid lines) and $v_{bc} = 0$ (dashed lines) at z = 20 (left) and z=10 (right). The suppression in the halo number density for the $v_{bc} = 2\times\sigma_{v_{bc}}$ case is largest at higher redshift with only 60\% as many haloes with masses $M_{DM}>10^5M_{\astrosun}$ at redshift 20 as in the zero-stream velocity run. With decreasing redshift, the suppression of the mass function also decreases, however remaining considerable at $z=10$, with 25\% fewer DM-primary objects with masses $M_{DM}>10^5M_{\astrosun}$ identified in the  $v_{bc} = 2\times\sigma_{v_{bc}}$ simulation. The suppression of the halo number density decreases with increasing halo mass and it is only on the order of 8\% for DM-primary objects with $M_{DM}>10^7M_{\astrosun}$, at redshift 10. These estimates are computed from the $1.4$ \Mpch\ simulation, but the results from the $700$ \kpch\ and the $1.12$ \Mpch\ are very similar. Notice that the $2.8$ \Mpch\ simulation overestimates the suppression due to the stream velocity effect. Since this simulation is performed at the coherence boundary of the stream velocity effect, the results obtained from these simulations regarding the smallest haloes cannot be fully trusted. Yet, we find that the halo mass function does converge at higher halo masses with those obtained from the higher resolution runs and, hence, we can safely use the $2.8$ \Mpch\ runs to infer the properties of haloes with masses larger than $10^7 - 10^8 M_{\astrosun}$.

\subsection{Suppression of gas fraction in haloes}\label{sec:DMpGASs}

In addition to the suppression of the halo abundance, previous studies have  shown that accounting for a non-zero stream velocity also impacts the gas content of dark matter haloes, with strong implications for the onset of reionization and the formation of the first stars \citep{Maio:2010qi, Greif:2011iv, O'Leary:2012rs, Fialkov:2011iw, Naoz:2011if, Naoz:2012fr, Tseliakhovich:2010yw, Dalal:2010yt, Richardson:2013uqa}. The non-zero stream velocity acts as an additional pressure term, increasing the minimum mass required for a dark matter halo to be able to accrete and retain gas. Furthermore, the gas density and temperature profiles are also affected, exhibiting a suppression of a factor of a few in the central parts of the halo \citep{O'Leary:2012rs}. Compared to numerical simulations consistently accounting for the impact of the non-zero stream velocity of the baryons that have been run in the past, here we systematically explore the impact of stream velocity on the gas content of haloes up to $10^9 M_{\astrosun}$, finding that the gas depletion is still considerable in haloes up to $M<10^8 M_{\astrosun}$.

\begin{figure*}
\centering
\begin{minipage}[b]{.4\textwidth}
\centering
\includegraphics[scale=.4]{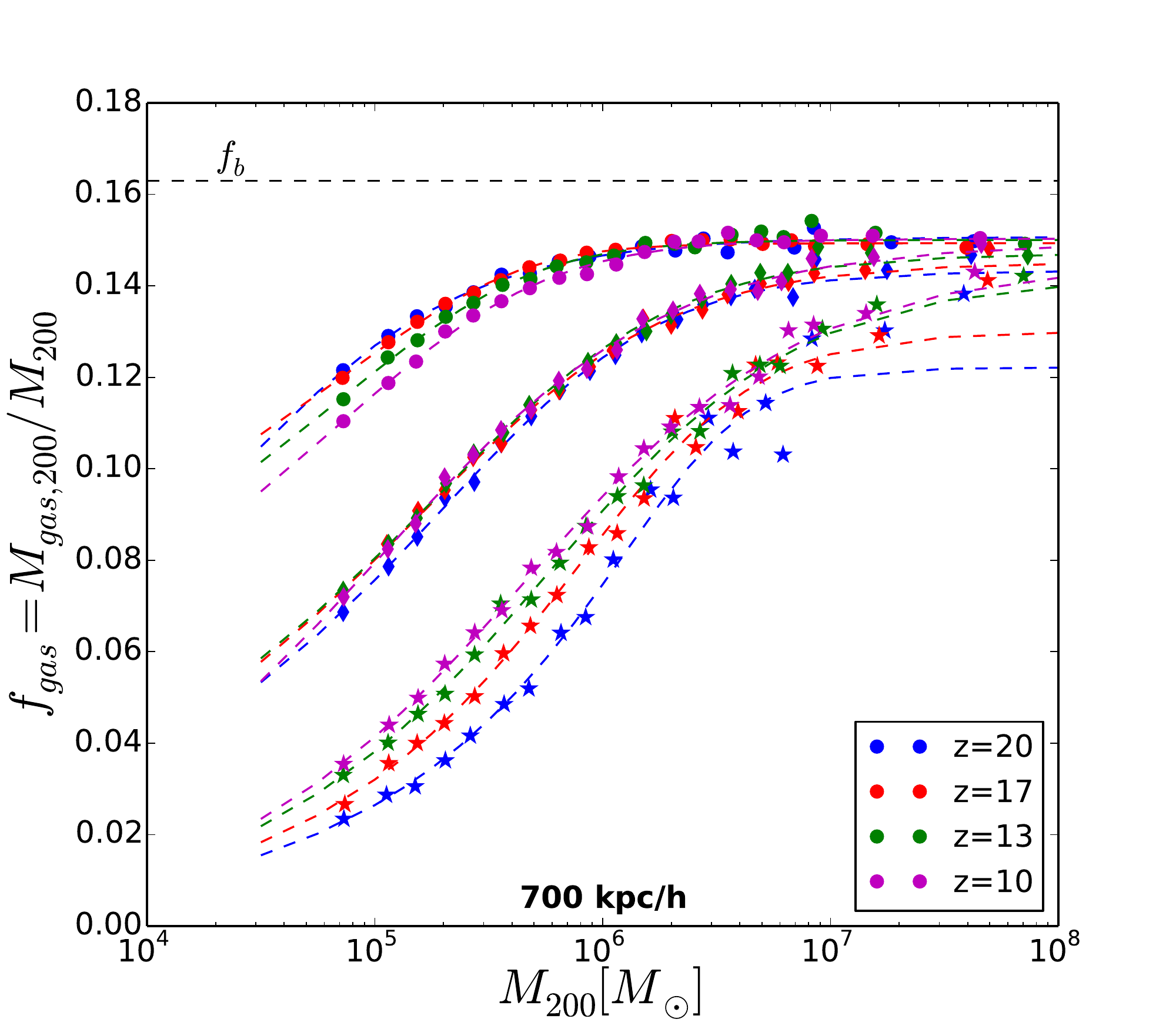}
\end{minipage}\qquad
\begin{minipage}[b]{.4\textwidth}
\centering
\includegraphics[scale=.4]{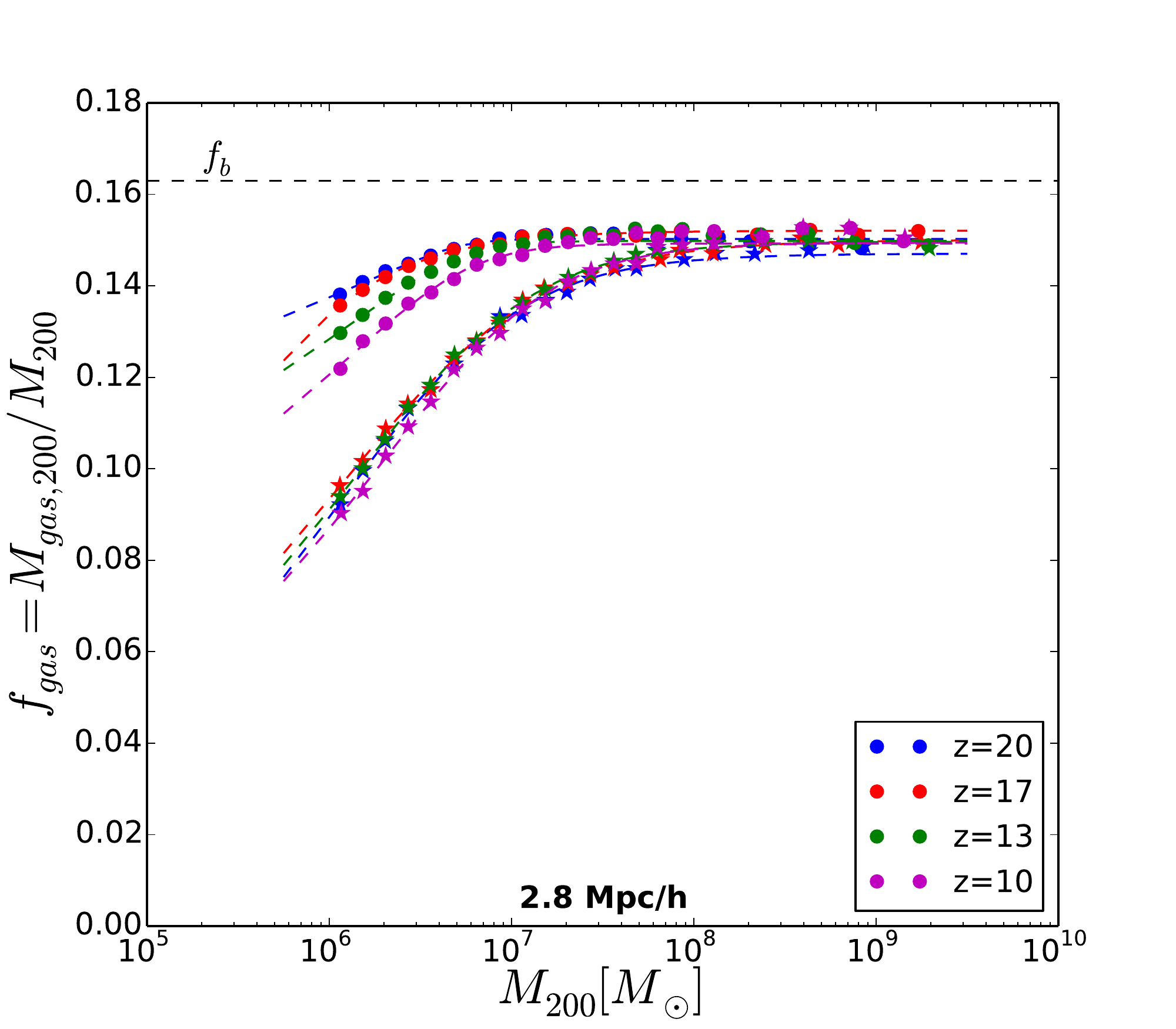}
\end{minipage}\qquad
\begin{minipage}[b]{.4\textwidth}
\centering
\includegraphics[scale=.4]{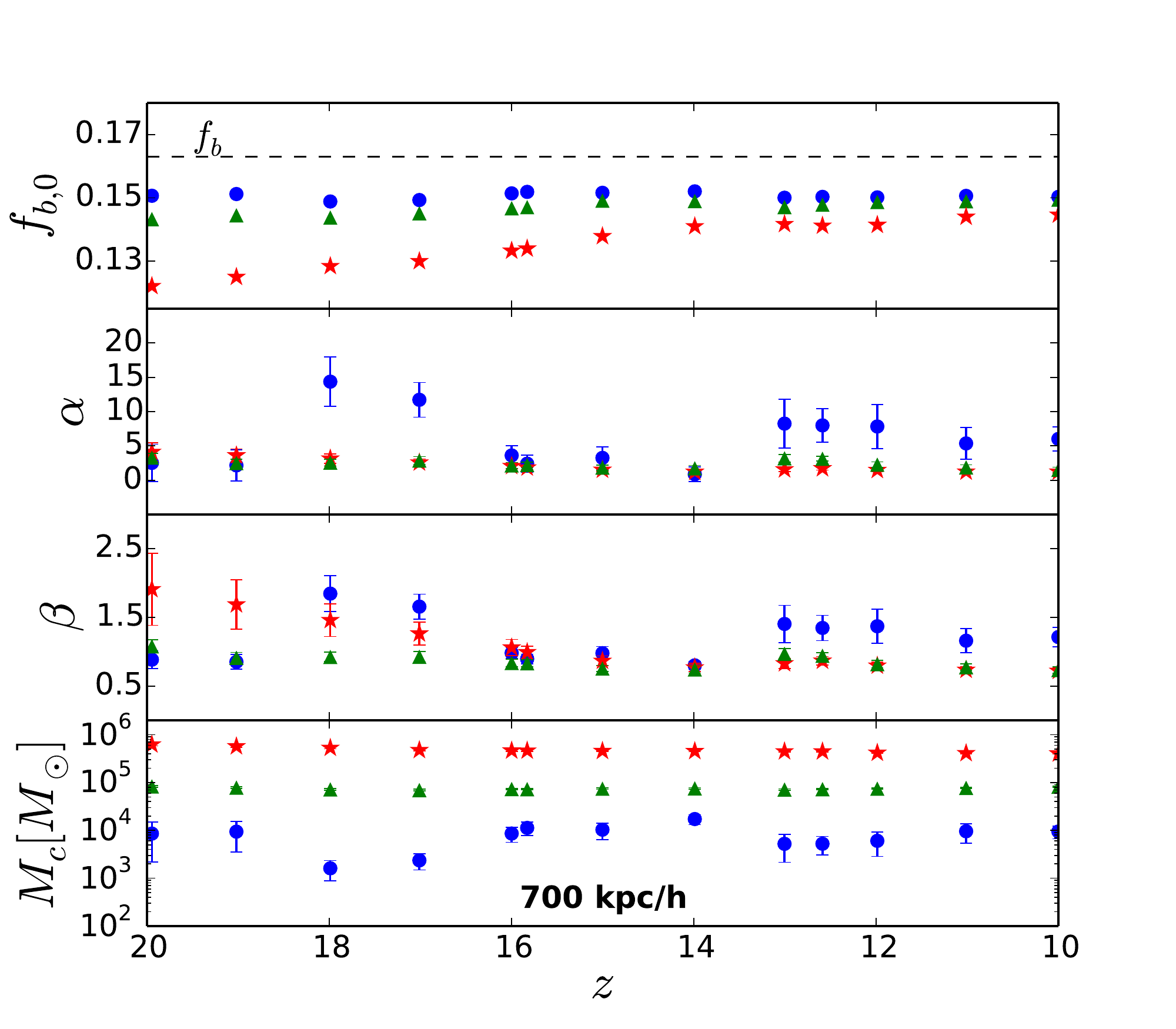}
\end{minipage}\qquad
\begin{minipage}[b]{.36\textwidth}
\centering
\includegraphics[scale=.4]{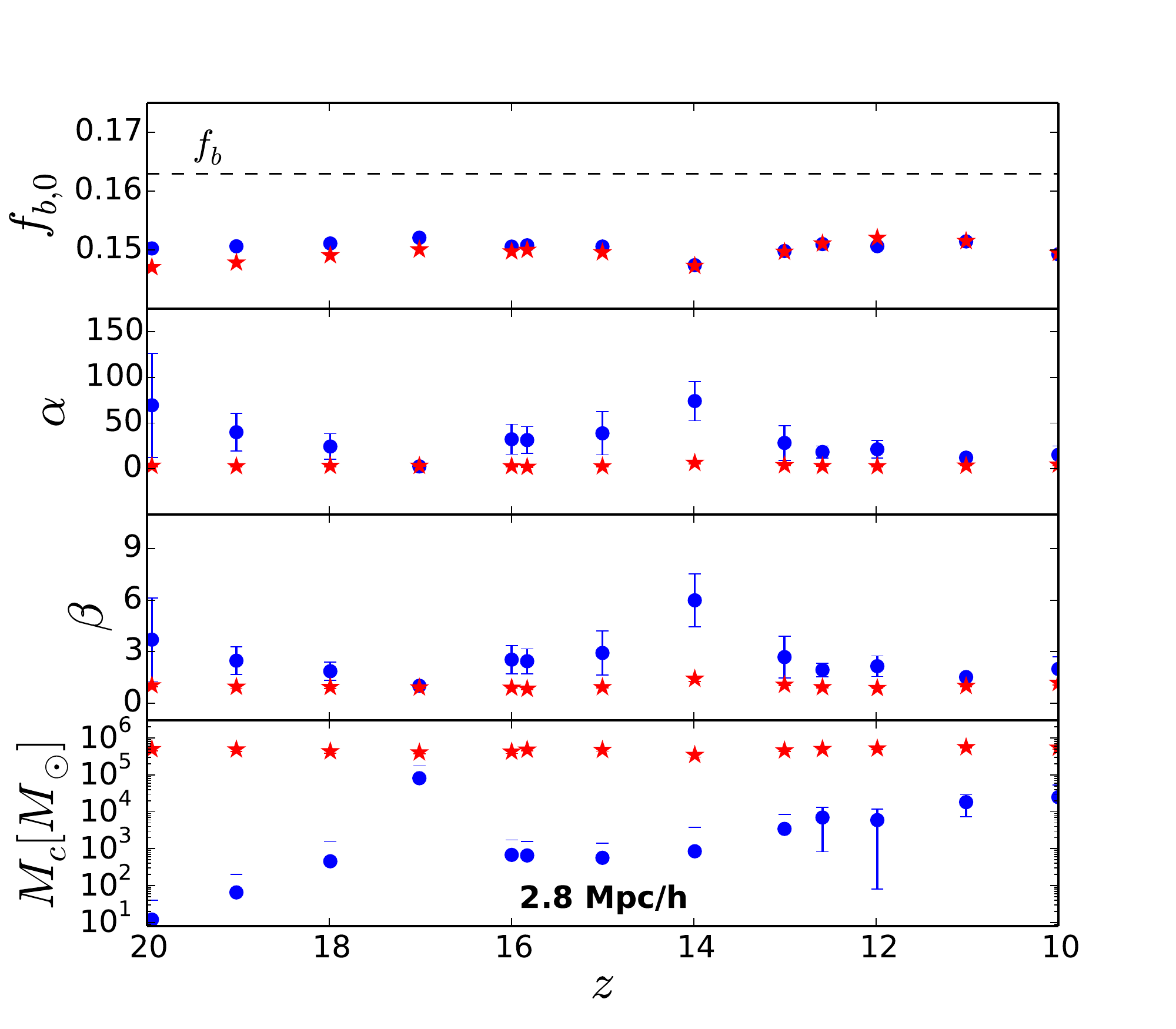}
\end{minipage}
\qquad\caption{Top: Average gas fraction inside DM-primary, Gas-secondary 
objects of a given total mass, together with the best fit functional form from 
Eq \ref{eq:naoz}. Circles represent data points from the $v_{bc} = 0$ runs, 
diamonds from the $v_{bc} = \sigma_{v_{bc}}$ run and stars from the $v_{bc} = 
2\times\sigma_{v_{bc}}$ runs. Bottom: Best fit parameters for the curves shown 
in top panels. Left column show results from the $700$ \kpch\ runs, while the 
right column shows results for the $2.8$ \Mpch\ runs.}
\label{fig:agvFgas3ParamOldFb0}
\end{figure*}

The top panels in Figure \ref{fig:rhofgasDMHalos} show the gas fraction as a function of the total mass of the halo, for DM-Primary, Gas-Secondary objects identified in our 700 \kpch\ runs. At redshift 20, in the high stream velocity case, a good fraction of the low mass haloes are virtually devoid of baryons, and, while this effect decreases with redshift, at $z=10$, the suppression in the gas fraction in haloes is still significant. The average gas fraction at redshift 20 inside haloes of $M \sim 10^6 M_{\astrosun} (10^7 M_{\astrosun})$ is $ \sim 40\%(15\%)$ lower in the $v_{bc} = 2\times\sigma_{v_{bc}}$ case than in the zero stream velocity case, while at redshift 10 the suppression in the gas fraction remains at $\sim30\%(10\%)$ level.  Note that our selection criterion for these objects imposes a minimum of 300 dark matter particles which effectively translates into a mass range of $2\times10^4M_{\astrosun} < M < 4\times10^7 M_{\astrosun}$ for haloes identified in our highest resolution run, at redshift 20. The larger box simulations explore haloes with masses up to few$\times10^9M_{\astrosun}$ and, as shown in Figure \ref{fig:rhofgasDMHalos1}, they, too, exhibit similar qualitative trends. A comparison between the fraction of haloes with a given gas fraction identified in the different runs shows that, in the case of zero stream velocity, the four different resolution runs are fairly well converged, indicating that already at $M\sim$ few$\times10^4M_{\astrosun}$ haloes are sufficiently massive to successfully accrete and retain gas (see bottom panels in Figure \ref{fig:rhofgasDMHalos}). On the other hand, the $v_{bc} = 2\times\sigma_{v_{bc}}$ runs do not exhibit such a convergence, indicating a much stronger dependence of gas fraction on halo mass. The highest resolution run, accessing lower mass objects, shows an abundance of haloes devoid of gas, while simulations testing the gas content of larger objects display a trend for the gas fraction distribution to approach the zero-stream velocity values.  At smaller redshifts, these differences decrease significantly, with the gas fraction distribution for the zero stream velocity case broadening, and the distribution for the non-zero stream velocity runs narrowing and shifting toward high gas fractions. This is partly due to the fact that, with decreasing redshift, larger haloes have time to form, while smaller haloes, in turn, have more time to accrete gas. 

\begin{figure*}
\centering
\begin{minipage}[b]{.4\textwidth}
\centering
\includegraphics[scale=.4]{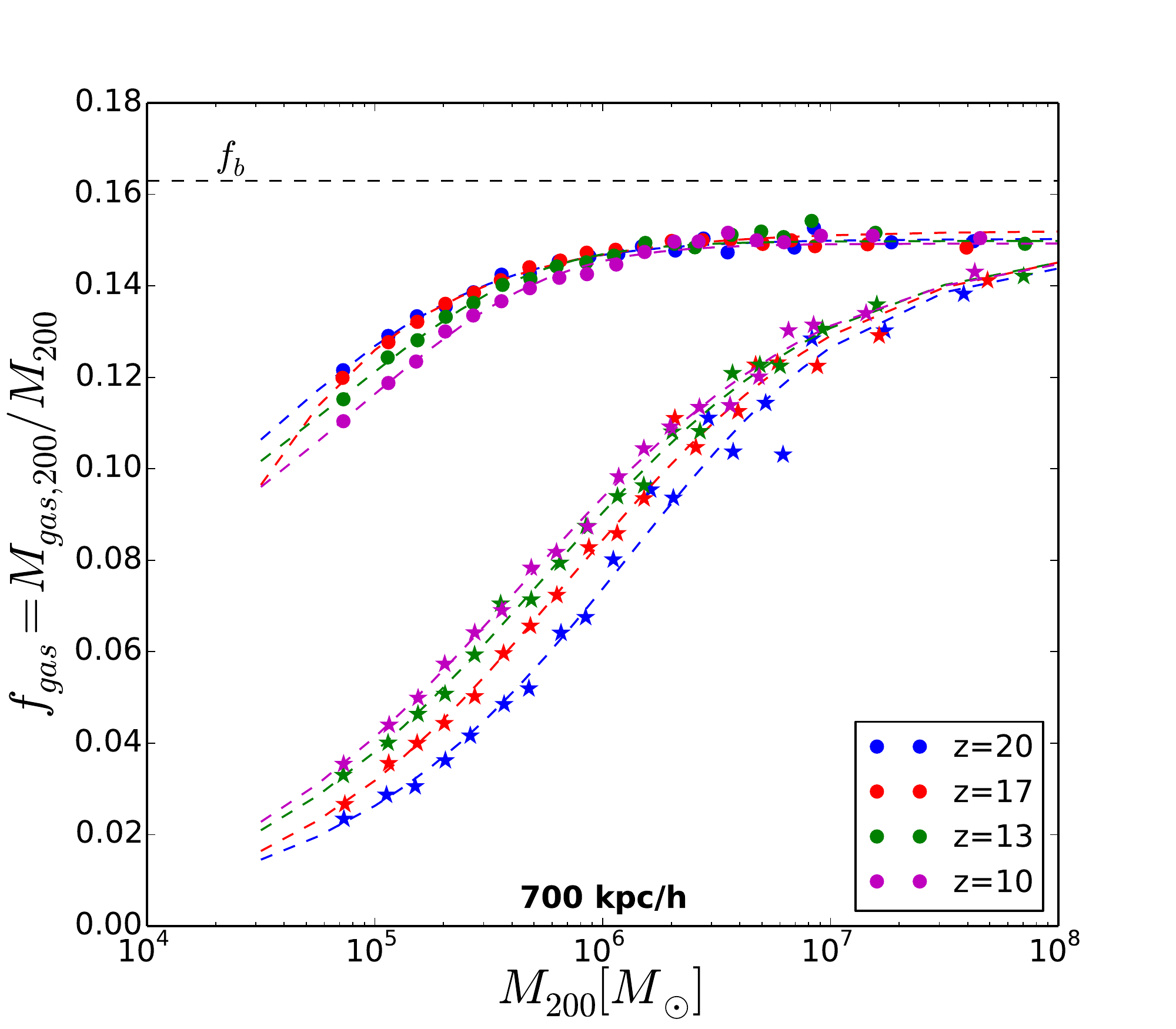}
\end{minipage}\qquad
\begin{minipage}[b]{.4\textwidth}
\centering
\includegraphics[scale=.4]{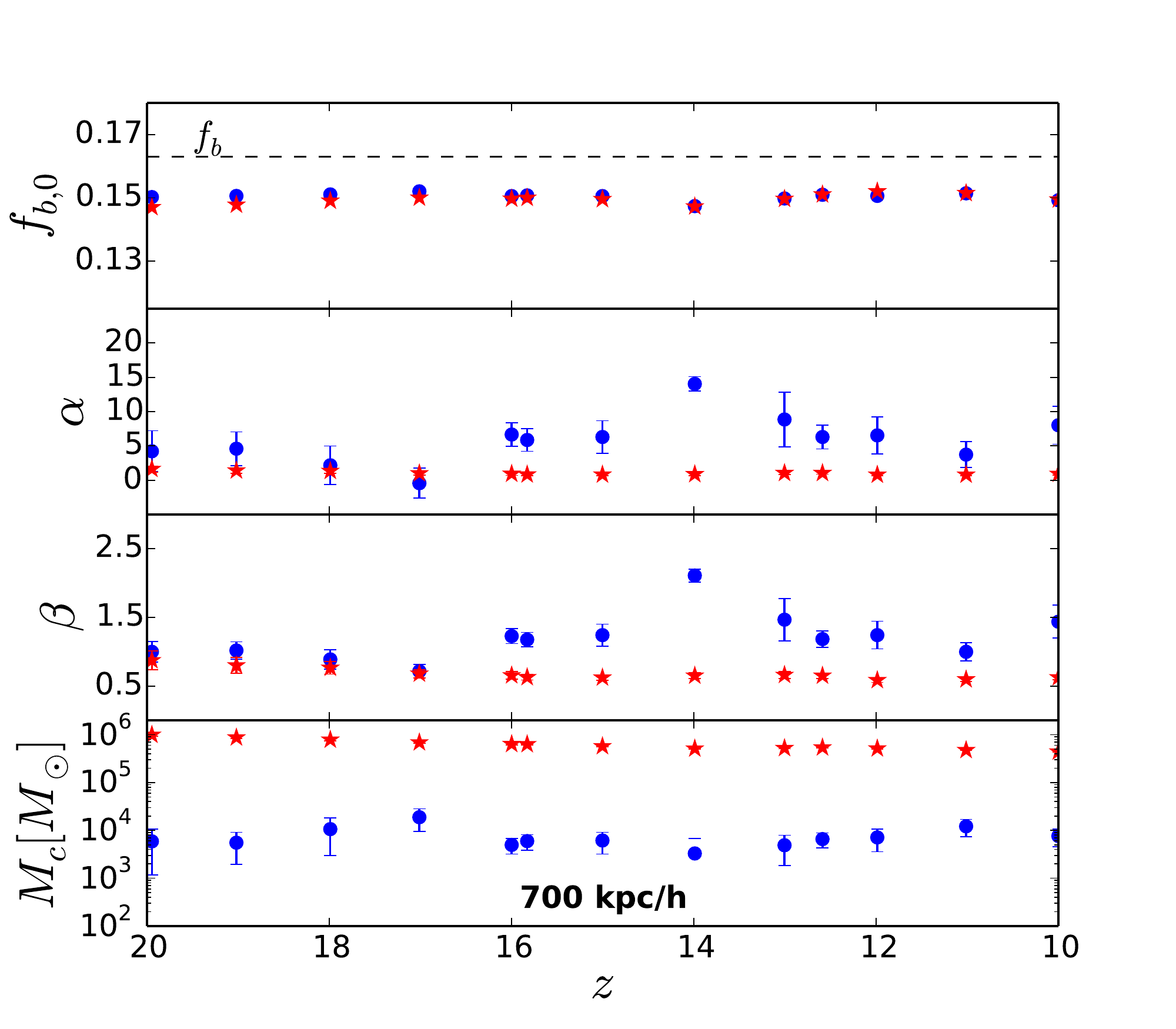}
\end{minipage}\qquad
\begin{minipage}[b]{.4\textwidth}
\centering
\includegraphics[scale=.4]{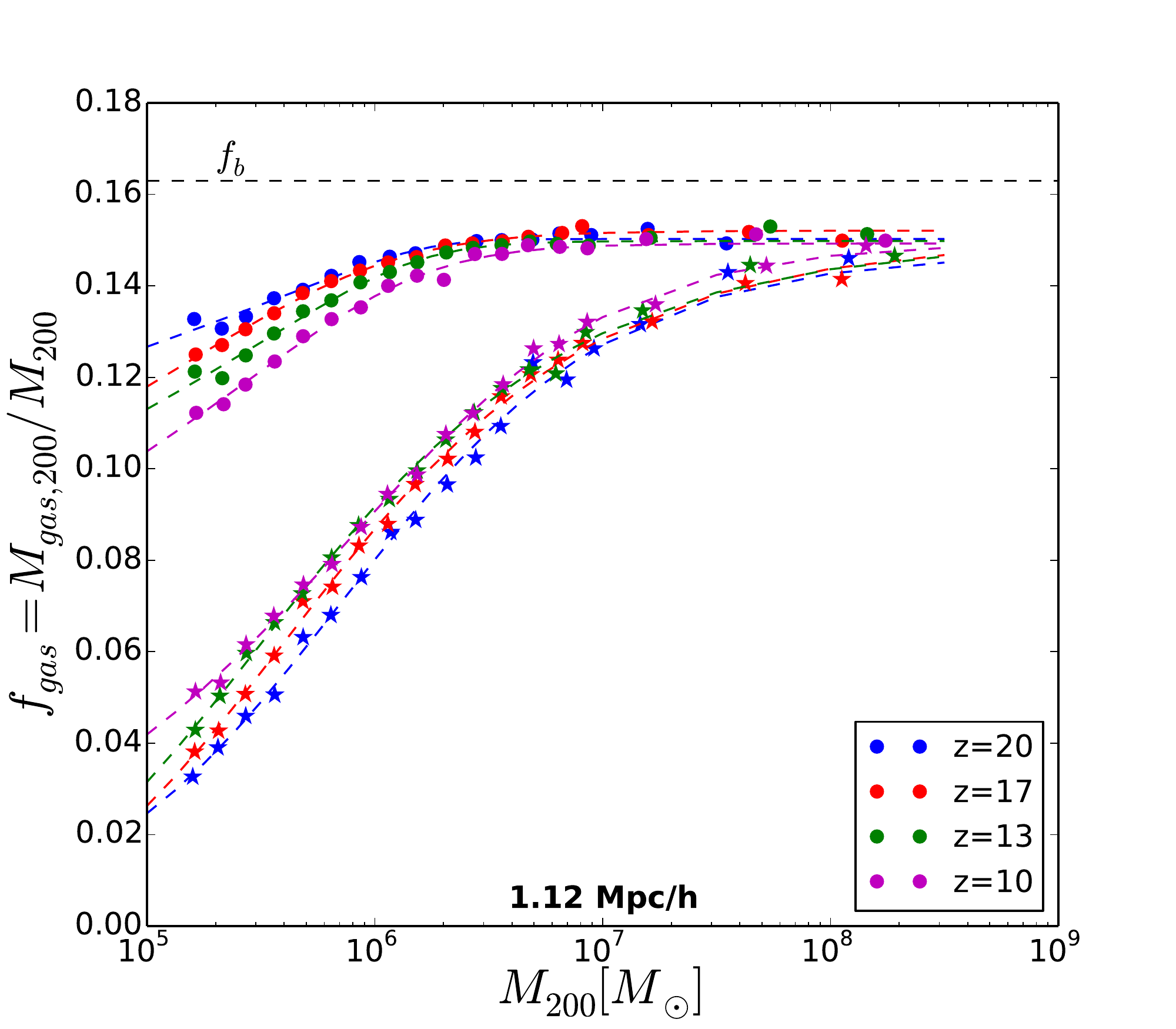}
\end{minipage}\qquad
\begin{minipage}[b]{.4\textwidth}
\centering
\includegraphics[scale=.4]{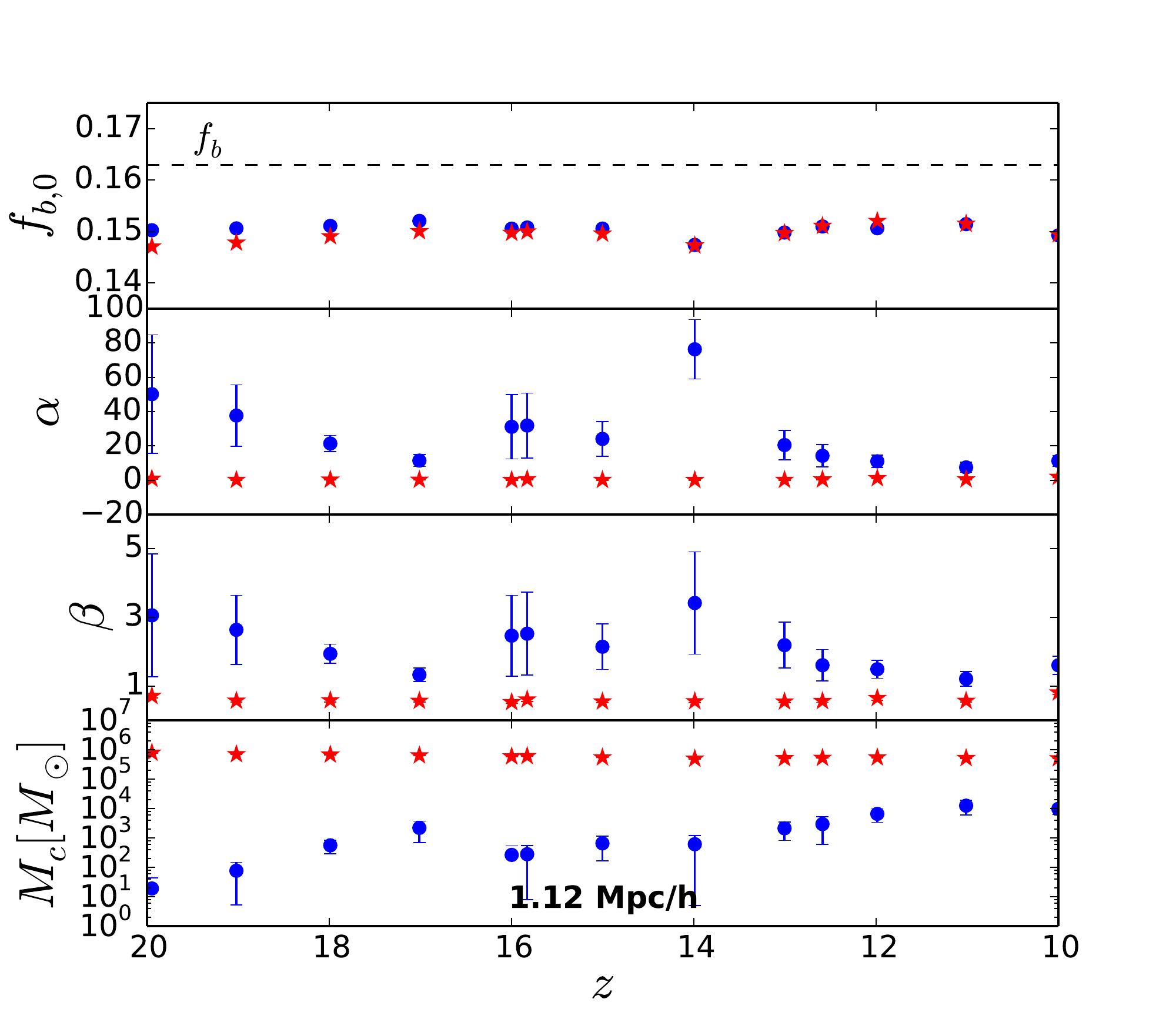}
\end{minipage}\qquad
\begin{minipage}[b]{.4\textwidth}
\centering
\includegraphics[scale=.4]{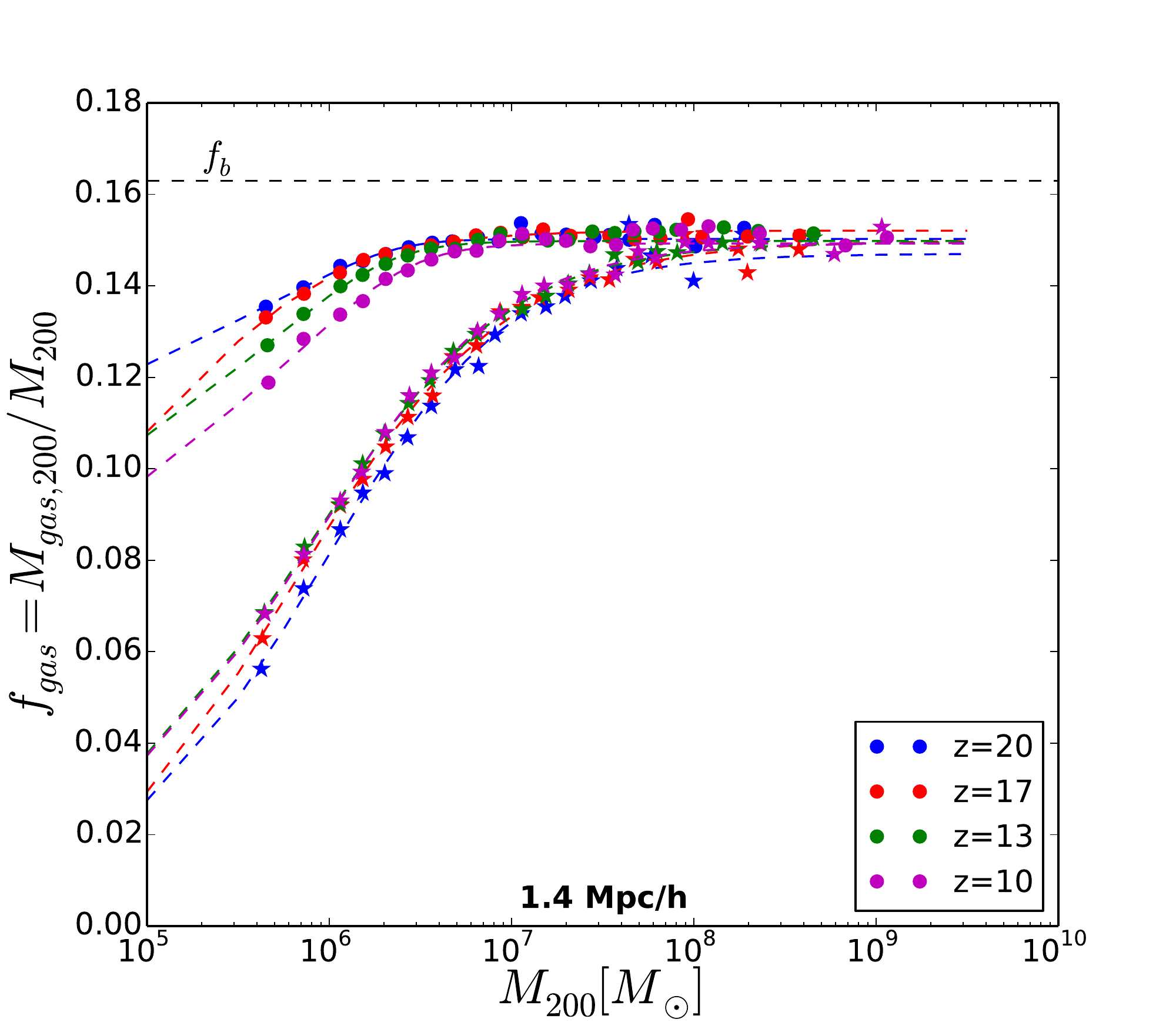}
\end{minipage}\qquad
\begin{minipage}[b]{.36\textwidth}
\centering
\includegraphics[scale=.4]{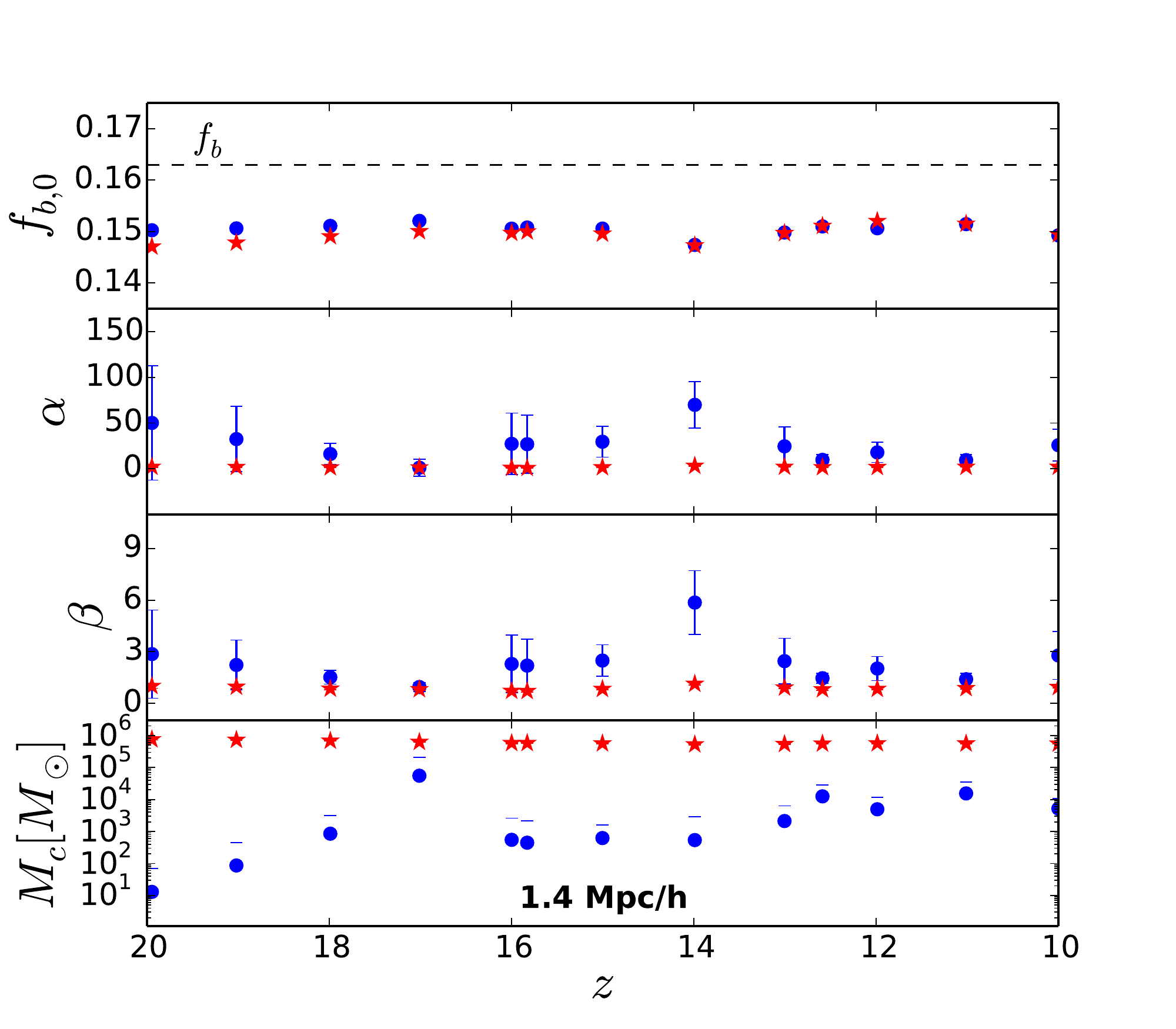}
\end{minipage}\qquad
\caption{Left: Average gas fraction inside DM-primary, Gas-secondary objects of a given total mass, together with the best fit functional form from Eq \ref{eq:naoz}, where $f_{b,0}$ has been computed using the information from the $2.8$ \Mpch\ runs. Circles represent data points from the $v_{bc} = 0$ runs, and stars from the $v_{bc} = 2\times\sigma_{v_{bc}}$ runs. Right: Best fit parameters for the curves shown in the panels on the left. This figure shows results from the $700$ \kpch\ , $1.12$ \Mpch\ , and $1.4$ \Mpch\ runs. The results for the $2.8$ \Mpch\ runs are, naturally, the same as in figure \ref{fig:agvFgas3ParamOldFb0}.}
\label{fig:agvFgas3ParamBest}
\end{figure*}

Small dark matter haloes have a suppressed gas fraction, regardless of the value of the stream velocity considered and only haloes with masses above a critical mass scale $M_C$ will have converged gas fraction distributions (e.g. \citealt{Gnedin:1997td, Gnedin:2000uj, Naoz:2006ye, Naoz:2009zu, Naoz:2010hg}). \cite{Naoz:2012fr} showed that that the functional dependence of the gas fraction on the mass of the halo, $M$ is given by: 
\begin{equation}
f_{gas} = f_{b,0} \left[ 1 + \left( 2^{\alpha/3}-1\right) \left(\frac{M_C}{M}\right)^\beta \right]^{-3/\alpha}
\label{eq:naoz}
\end{equation}
where $\beta$ and $\alpha$ are the slopes governing the steepness of the dependence of the gas fraction on the halo mass in the low-mass regime and the sharpness of the transition from the high-mass end to the low-mass end of the spectrum, respectively. Both of these parameters, as well as the critical mass scale $M_C$ are fit from simulation results. $f_{b,0}$ is the gas fraction in the high-mass end of the spectrum and it is defined as the gas fraction in the heaviest $5\%$ of the haloes realized \citep{Naoz:2012fr}. This definition of $f_{b,0}$ is based on the observation that haloes with masses above a certain threshold will have a similar gas fraction, in which case, as long as the most massive haloes in a simulation have masses above this threshold, their gas fraction will be $\sim f_{b,0}$. For our $v_{bc} = 0$ runs, the distribution of gas fraction in haloes is well converged between the four different resolution runs even though they access different halo mass ranges, indicating that the gas fractions in the heaviest haloes are $\sim f_{b,0}$ regardless of the box size of the run. However, as shown in the bottom panels of Figure \ref{fig:rhofgasDMHalos}, for the runs with $v_{bc} = 2\times\sigma_{v_{bc}}$ the gas fraction distributions are not converged between the different resolutions, indicating that gas fraction in the haloes analyzed is still very strongly dependent on the mass of the object. Therefore, it is likely that the heaviest haloes realized in at least some of these simulations are, in fact, not massive enough and contain a gas fraction significantly below $f_{b,0}$.

This fact is confirmed by inspecting the best fit of the dependence of the gas fraction on halo mass, as given by Equation \ref{eq:naoz}. As shown in Figure \ref{fig:agvFgas3ParamOldFb0}, the data from both the zero stream velocity and high stream velocity runs with box size $2.8$ \Mpch\ is well fitted by Equation \ref{eq:naoz}, indicating that the functional form captures well the shape of the dependence and the computed $f_{b,0}$ is a good estimate of the asymptotic value of the gas fraction in the high-mass limit. The characteristic mass scale, given by $M_C$, below which the gas fraction in haloes is significantly suppressed, is much lower in the zero stream velocity case, while for $v_{bc} = 2 \sigma_{v_{bc}}$ it is on the order of few$\times10^5 M_{\astrosun}$ at all redshifts in the $20<z<10$ range. The value $f_{b,0}$ to which the gas fraction asymptotes in the high mass limit is, however, independent of the value of the stream velocity, in agreement with previous studies which showed that the suppression in the gas fraction due to a non-zero stream velocity is present for haloes with masses $M \lesssim 10^7-10^8 M_{\astrosun}$ (see also \citealt{Maio:2010qi, Greif:2011iv, O'Leary:2012rs, Fialkov:2011iw, Naoz:2011if, Naoz:2012fr, Tseliakhovich:2010yw, Dalal:2010yt, Richardson:2013uqa}).

In the case of simulations with non-zero stream velocity and a box size less than $2.8$ \Mpch\ , Equation \ref{eq:naoz} provides a significantly worse fit for the gas fraction as a function of halo mass, particularly in the high mass region, where the fitted function asymptotes at a significantly lower gas fraction than that observed in the simulation. The most striking example of this misfit of the high-mass end of the spectrum is realized for the $700$ \kpch\ simulations and is shown in Figure \ref{fig:agvFgas3ParamOldFb0}. This discrepancy is due to the fact that $f_{b,0}$ is not a fitted parameter, but rather it is given as input to the function, and computed as the gas fraction in the top 5\% heaviest haloes. However, in the smaller boxes, the 5\% heaviest haloes are still too light to reproduce the true high mass limit of the gas fraction distribution for the non-zero stream velocity case. Therefore, in order to produce a meaningful fit of the distribution, $f_{b,0}$ should either be considered as another free parameter to fit together with $\alpha$, $\beta$ and $M_C$ or it should be computed from a simulation in which the high mass limit is, in fact, realized. Here we opt for the latter solution, and adopt the value for $f_{b,0}$ computed from the $2.8$ \Mpch\ box. The new best fit functional form for the $700$ \kpch\ data is shown in Figure \ref{fig:agvFgas3ParamBest}. The fit of the non-zero stream velocity simulation is much improved and now the behavior in the high mass region of the spectrum is appropriately captured. 

We should note that the values of $f_{b,0}$ computed from the different resolution runs for the $v_{bc} = 0$ case are, in fact, well converged and, hence, for the zero-stream velocity runs the quality of the fit of Equation \ref{eq:naoz} is roughly independent of box size of the run used to compute the high-mass limit of the gas fraction. This observation is in agreement with the fact that, for the $v_{bc} = 0$ runs, the best fit characteristic mass (Figures \ref{fig:agvFgas3ParamOldFb0} and \ref{fig:agvFgas3ParamBest}) below which the gas fraction in haloes is significantly suppressed is more than 3 orders of magnitude lower than the highest halo mass realized even in the simulation with the smallest box size. Finally, we find that the average gas fraction in haloes with masses $M \lesssim 10^6 M_{\astrosun}$ is significantly higher in our high stream velocity $2.8$ \Mpch\ simulation, than that recovered in similar mass haloes realized in our non-zero stream velocity runs of smaller boxes. We attribute this discrepancy  to the fact that the $2.8$ \Mpch\ box size lies at the edge of the coherence scale of the stream velocity and, as discussed in the context of the halo mass functions, it does not accurately recover the low halo mass behavior displayed in the other runs. However, the largest simulation does converge to the expected results in the high-mass end of the spectrum and, therefore, we can rely on the $f_{b,0}$ value computed from this run. Furthermore, as shown in Figure \ref{fig:agvFgas3ParamBest} we find that the average gas fractions as a function of halo mass  recovered from the $700$ \kpch\ , $1.12$ \Mpch\ , and $1.4$ \Mpch\ runs agree well over a large range of masses.
We note that $f_{b,0}$ is lower than $f_{b}$ even in the zero-stream velocity case because the baryon over-densities have smoother initial conditions than the dark matter component and, therefore, it takes the gas a timescale longer than that explored in our simulation to fully sink into the DM potential wells, in agreement with the analytical predictions in \cite{Naoz:2006ye}.

\begin{figure}
\centering
\includegraphics[scale=.4]{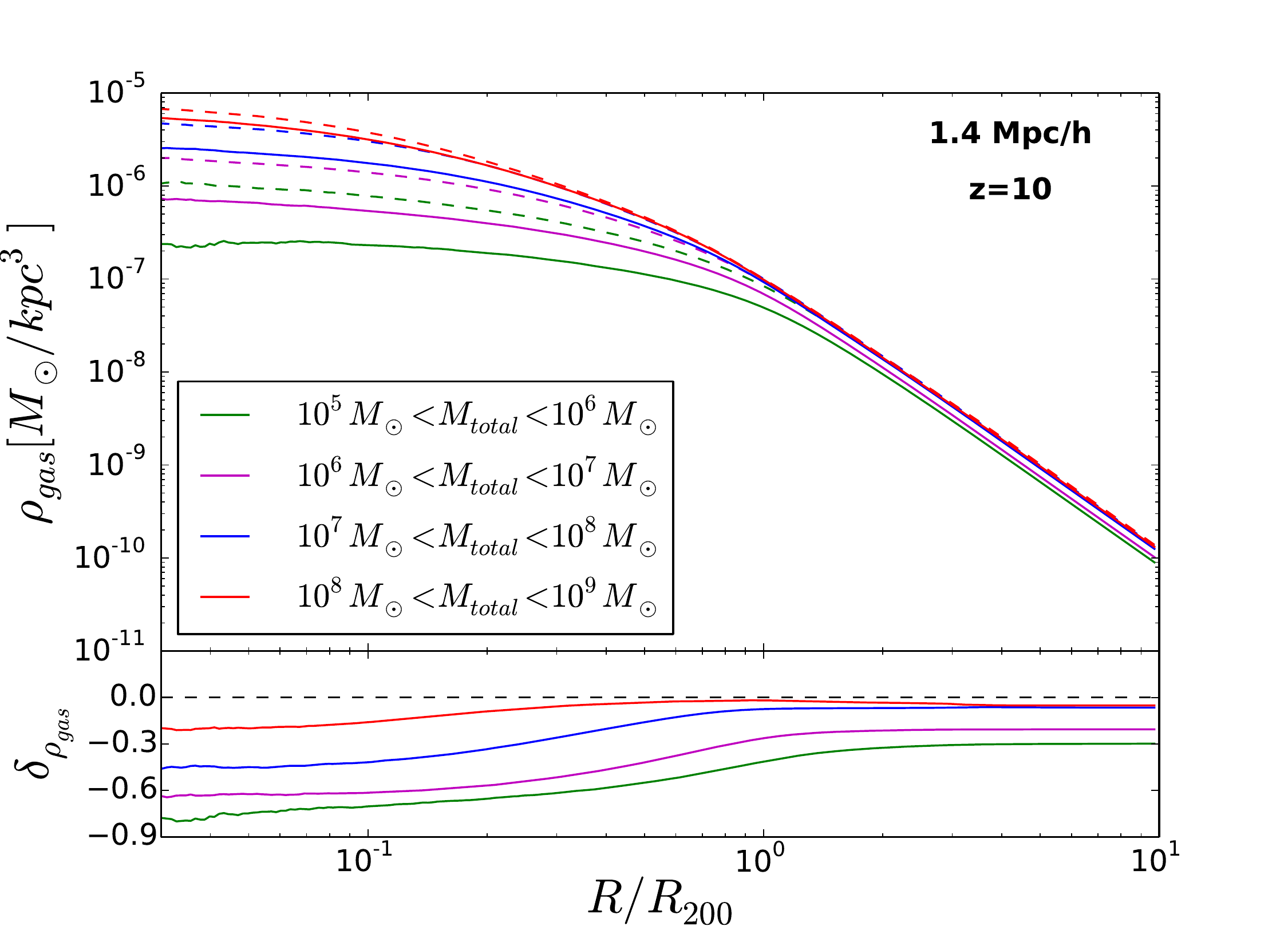}
\caption{Average gas density profiles for the DM-primary, Gas-secondary objects 
identified in the $1.4$ \Mpch\ simulations with $v_{bc} = 0$ (dashed lines) and 
$v_{bc} = 2\times\sigma_{v_{bc}}$ (solid lines), at redshift 10. The bottom 
panel shows the average value of the over-density identified in the high stream 
velocity case $\ds\left( \delta_{\rho_{gas}} = (\rho_{gas,2\sigma} - 
\rho_{gas,0})/\rho_{gas,0}\right)$.}
\label{fig:agvRhoGasMassBin}
\end{figure}

\begin{figure*}
\centering
\begin{minipage}[b]{.4\textwidth}
\centering
\includegraphics[scale=.38]{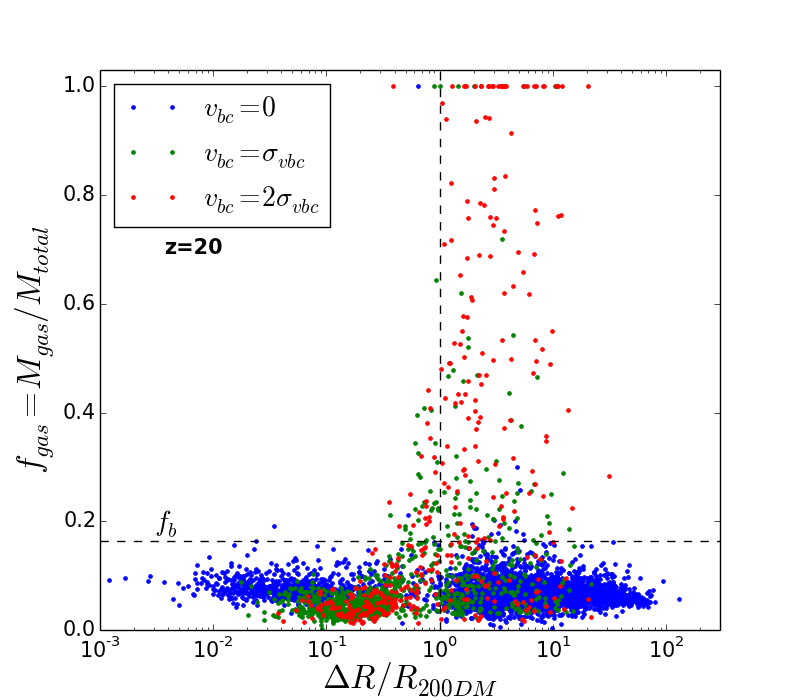}
\end{minipage}\qquad
\begin{minipage}[b]{.4\textwidth}
\centering
\includegraphics[scale=.38]{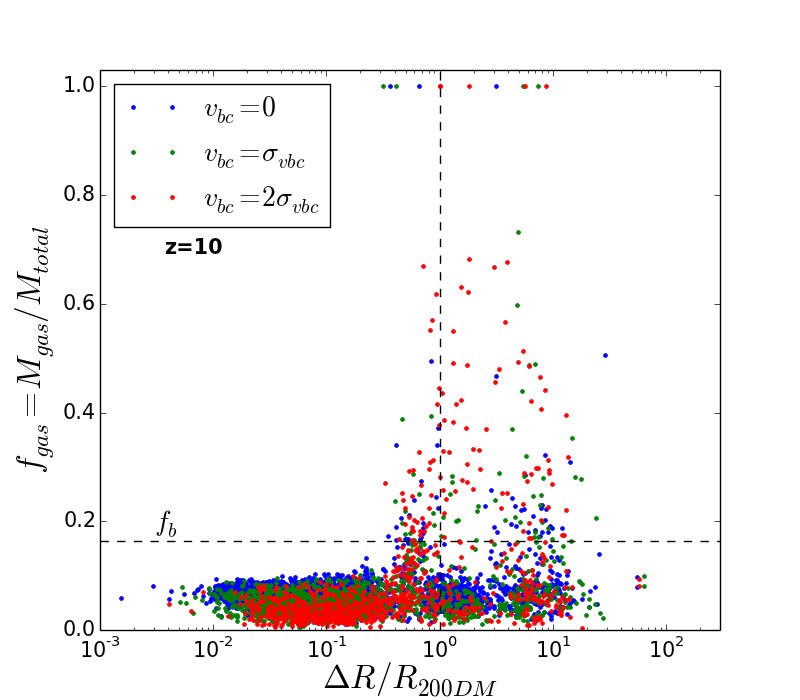}
\end{minipage}
\begin{minipage}[b]{.4\textwidth}
\centering
\includegraphics[scale=.38]{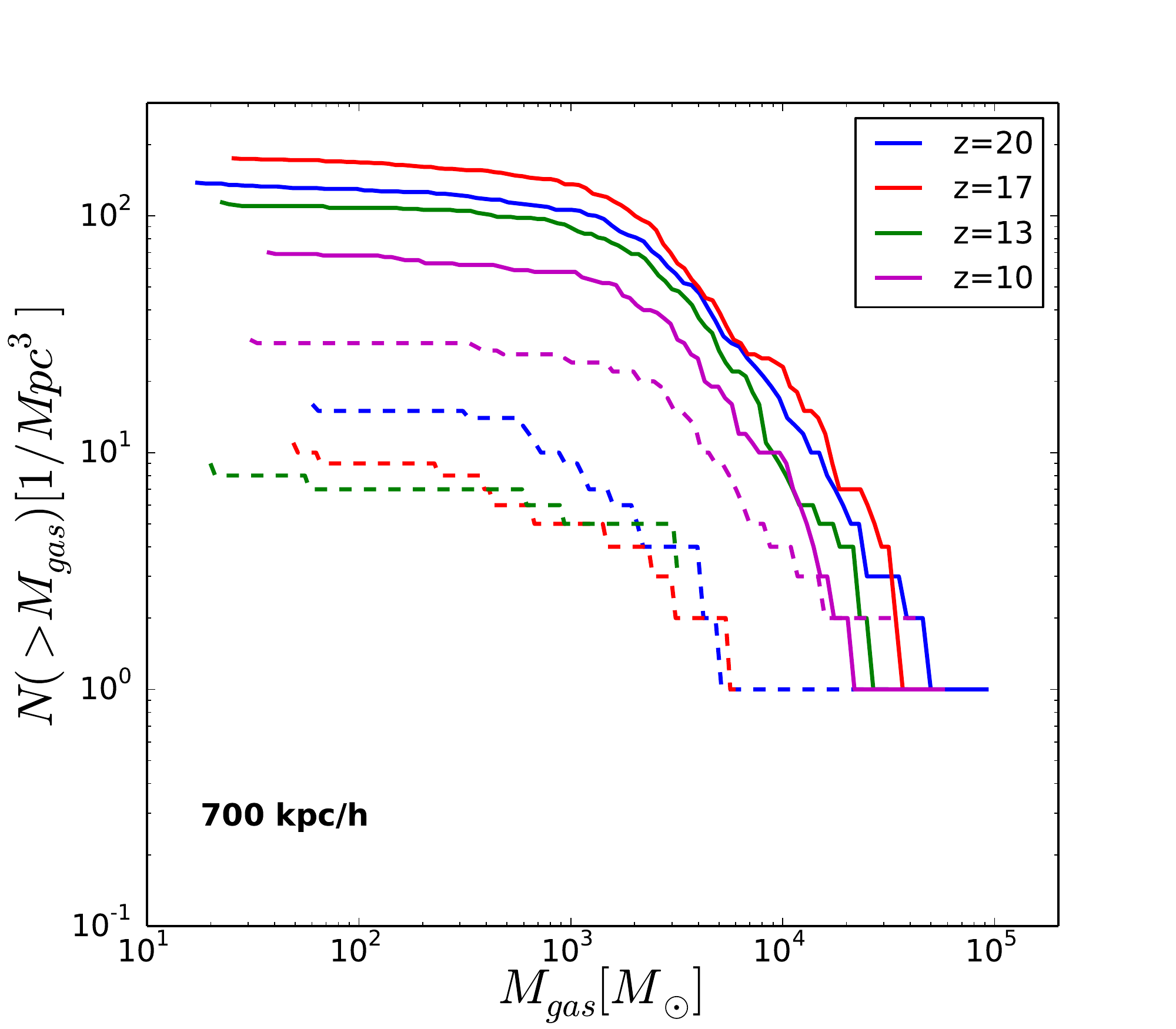}
\end{minipage}\qquad
\begin{minipage}[b]{.39\textwidth}
\centering
\includegraphics[scale=.38]{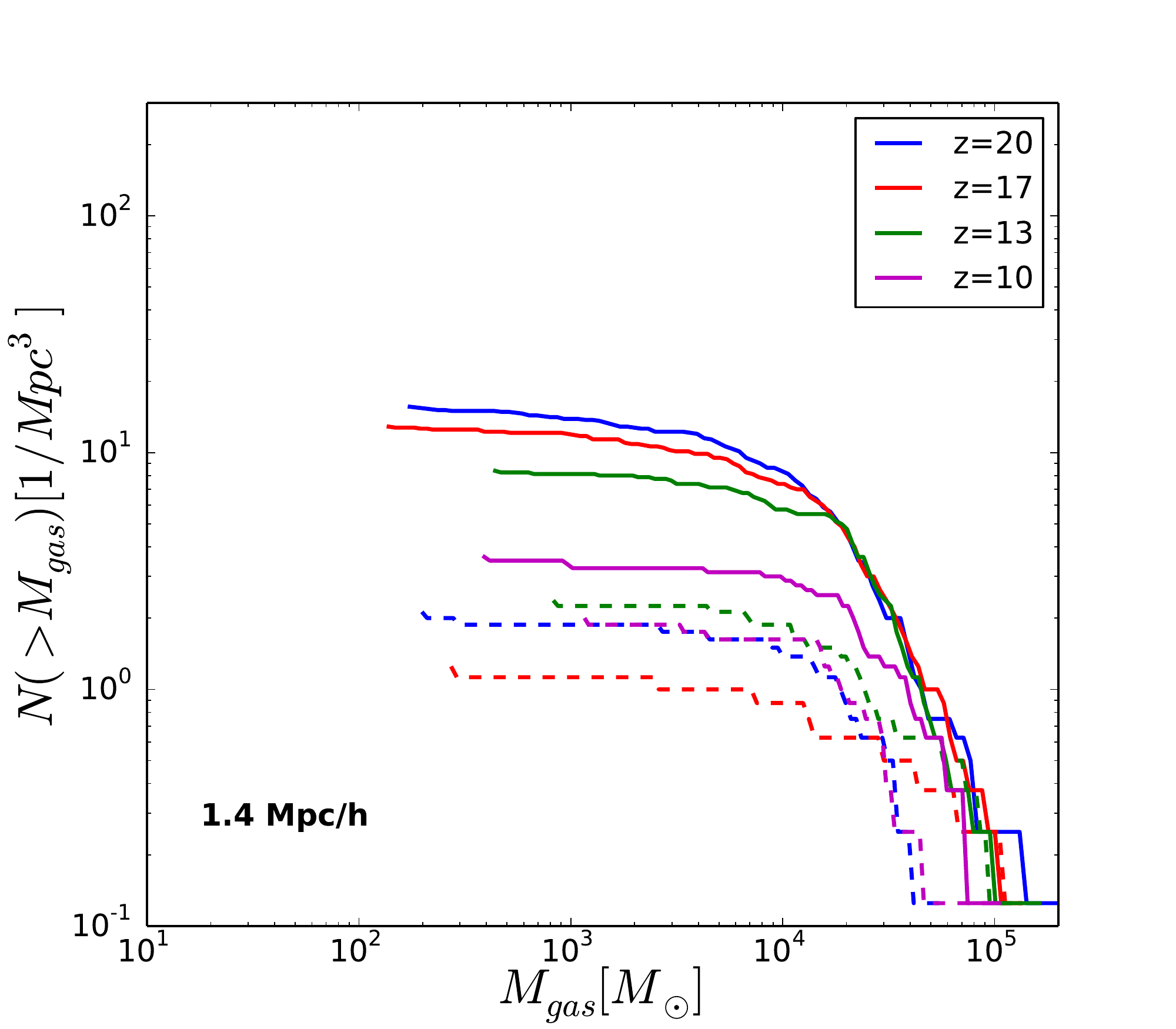}
\end{minipage}
\caption{Top: Gas fraction as a function of the distance to the closest dark matter halo for the Gas-primary objects identified in the $700$ \kpch\ simulations. All quantities are computed inside the optimal ellipsoid, as defined in Section \ref{sec:halodef}. Bottom: Halo mass function for the gas abundant gas-primary structures sitting outside the virial radius of the closest dark matter halo. The solid lines are computed from the high-stream velocity runs, while the dashed lines, from the zero-$v_{bc}$ simulations (see Figure \ref{fig:fgasmgasdeltar1} for results from the other simulation boxes).}
\label{fig:fgasmgasdeltar}
\end{figure*}

Interestingly, the non-zero stream velocity impacts not only the overall gas fraction inside haloes, but also the shape of the distribution of the gas density inside the dark matter haloes. Figure \ref{fig:agvRhoGasMassBin} shows the average gas density profiles computed for DM-primary, Gas-secondary objects with total masses split into mass bins as indicated in the legend. As expected from our analysis of the total gas fraction in these objects, for smaller haloes, the average gas density inside the virial radius of the DM haloes identified in the high stream velocity simulations plateaus to a fraction of that recovered inside the DM haloes from the $v_{bc} = 0$ simulations. On the other hand, for larger haloes, the gas fraction is virtually independent of the value of the stream velocity. However, for all haloes, regardless of how the total gas fraction compares to that recovered in the $v_{bc} = 0$ simulations, the gas density in the center region of the objects is substantially lower in the high stream velocity case. For haloes as massive as $\sim 10^9M_{\astrosun}$ at redshifts as low as 10, the suppression in the gas density in the inner regions of the halo is on the order of 25\%, indicating that for a non-zero stream velocity the accreted gas tends to linger in the outer regions of the haloes, in agreement with the qualitative results of \cite{O'Leary:2012rs}. This observation may be partially due to the fact that our simulations do not account for radiative cooling processes and only follow the gas evolution adiabatically. For that reason, the gas that has collapsed into dark matter haloes cannot efficiently cool and the density inside the inner parts of the halo asymptotes to a value determined by the equilibrium between pressure and gravity. Since a non-zero stream velocity effectively acts as an additional pressure term, the expected density inside dark matter haloes is naturally lower than that recovered in the zero stream velocity case.


\subsection{Gas-Primary objects}\label{sec:gasFrac_baryClumps}\label{sec:GASp}

\begin{figure*}
\centering
\begin{minipage}[b]{.29\textwidth}
\centering
\includegraphics[scale=1.2]{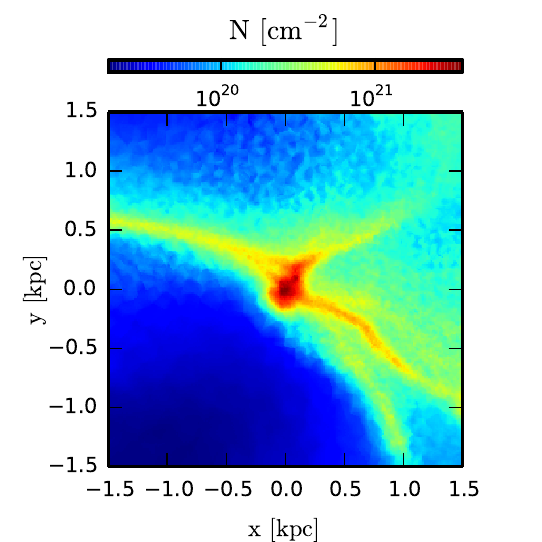}
\end{minipage}\qquad
\begin{minipage}[b]{.29\textwidth}
\centering
\includegraphics[scale=1.2]{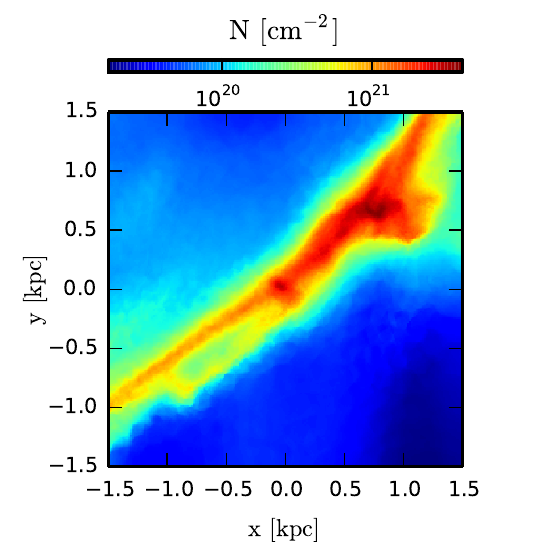}
\end{minipage}\qquad
\begin{minipage}[b]{.29\textwidth}
\centering
\includegraphics[scale=1.2]{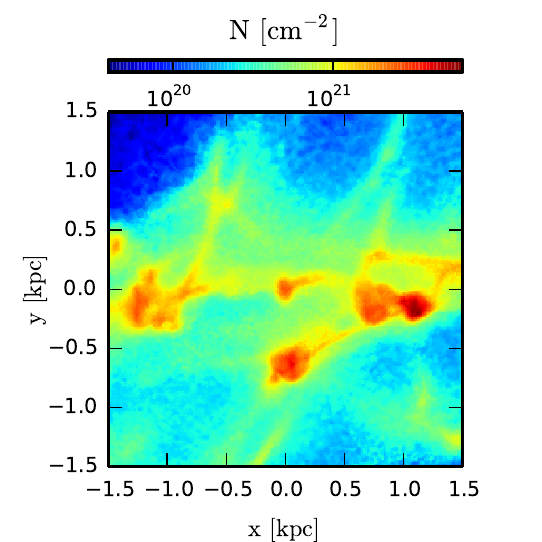}
\end{minipage}\qquad
\qquad\caption{Gas density projections for three Gas-Primary objects identified in the high stream velocity, 1.4 \Mpch\ simulation at redshift 10. The objects are centered at coordinates (0, 0) in the figures above, contain a gas fraction larger than the cosmic baryonic fraction inside the tightly fitted ellipsoids (not shown) and are located outside the virial radius of the closest DM halo.}
\label{fig:3dEllipsoids_DensityProfiles}
\end{figure*}

The analysis of the abundance and gas content of dark matter haloes in our simulations has shown that an increase in the stream velocity correlates with a suppression in the halo mass function, as well as with a decrease in the gas fraction of dark matter haloes with masses $< 10^7 M_{\astrosun}$. Given that all runs contain the same cosmic baryonic fraction,  a natural question that arises is, then, what happened to the gas that did not end up inside dark matter haloes. In a recent work, \cite{Naoz:2014bqa} showed that a non-zero relative 
velocity induces an offset between the baryonic over-densities and the DM ones. For baryonic modes with $M_{\rm b}\lsim $ few $ \times 10^6$~M$_\odot$, the baryon over-densities can reach non-linearity well outside the DM halo virial radius. \cite{Naoz:2014bqa} show that this process may  link these gas rich structures with the progenitors of globular clusters. Furthermore, the  gas deficient haloes, discussed in Section \ref{sec:DMpGASs} may be observed today as dark satellites. 

In order to identify and analyze the properties of gas-dominated objects which might form and survive outside dark matter haloes, we have defined a Gas-primary object, as described in Section \ref{sec:halodef}. In our analysis, we consider only FOF groups with more than 100 gas cells\footnote{We chose to lower the default number of required particles in the analyzed FOF groups from 300 to 100 in order to access a wider mass range of Gas-Primary objects.}, but place no constraints on the number of particles inside the best fit ellipsoid. However, due to the fact that the best fit ellipsoid is required to contain at least 80\% of the number of particles in the FOF group, this effectively places a lower mass bound on the mass of the Gas-Primary structures identified of $M_{gas} \geq 3.5\times10^3M_{\astrosun}, 9.5\times10^3M_{\astrosun}, 2.9\times10^4M_{\astrosun}, 6.9\times10^4M_{\astrosun}$ for the 700 \kpch\ , 1.12 \Mpch\ , 1.4 \Mpch\ and 2.8 \Mpch\ simulations, respectively. 

We find that a large fraction of the Gas-primary objects identified in our simulations are simply the gas component of the DM-primary, Gas-secondary objects discussed in Section \ref{sec:DMpGASs}. However we remind that the DM-primary, Gas-secondary objects are centered on the center of mass of the dark matter + gas system, for which spherical symmetry is assumed, and their properties are studied inside of the virial radius of the halo. Gas-primary objects are, on the other hand, centered on the center of mass of the gas, spherical symmetry is not required and their properties are studied inside a tightly fitting ellipsoid surrounding the gas particles. For that reason, the gas-primary objects which are also part of a DM-primary, Gas-secondary object are generally displaced with respect to the center of mass of the latter and exhibit, on average, a larger gas fraction than the one found for the the corresponding DM-primary, Gas-secondary object, discussed in Section \ref{sec:DMpGASs}.

In addition to the Gas-primary objects which can be identified as belonging to a dark matter halo, we also recover a number of objects which appear to reside {\it outside} of the virial radius of any DM-primary halo. For all Gas-primary objects constructed in our simulations, we identify a closest DM-primary object containing more than 300 dark matter particles and we show in the top panels of Figure \ref{fig:fgasmgasdeltar} the gas fraction contained in the Gas-primary objects as a function of their distance to the closest dark matter halo for our highest resolution runs. Note that, as mentioned in Section \ref{sec:halodef}, while we use just the gas particle information to identify Gas-primary objects, the best fit ellipsoidal surfaces surrounding the object will generally enclose both gas as well as dark matter particles. Therefore, the gas fraction associated with a Gas-primary object is computed as the gas mass divided by the total mass contained inside the best fit ellipsoids around the center of mass of the object. 

\begin{figure*}
\centering
\begin{minipage}[b]{.4\textwidth}
\centering
\includegraphics[scale=.36]{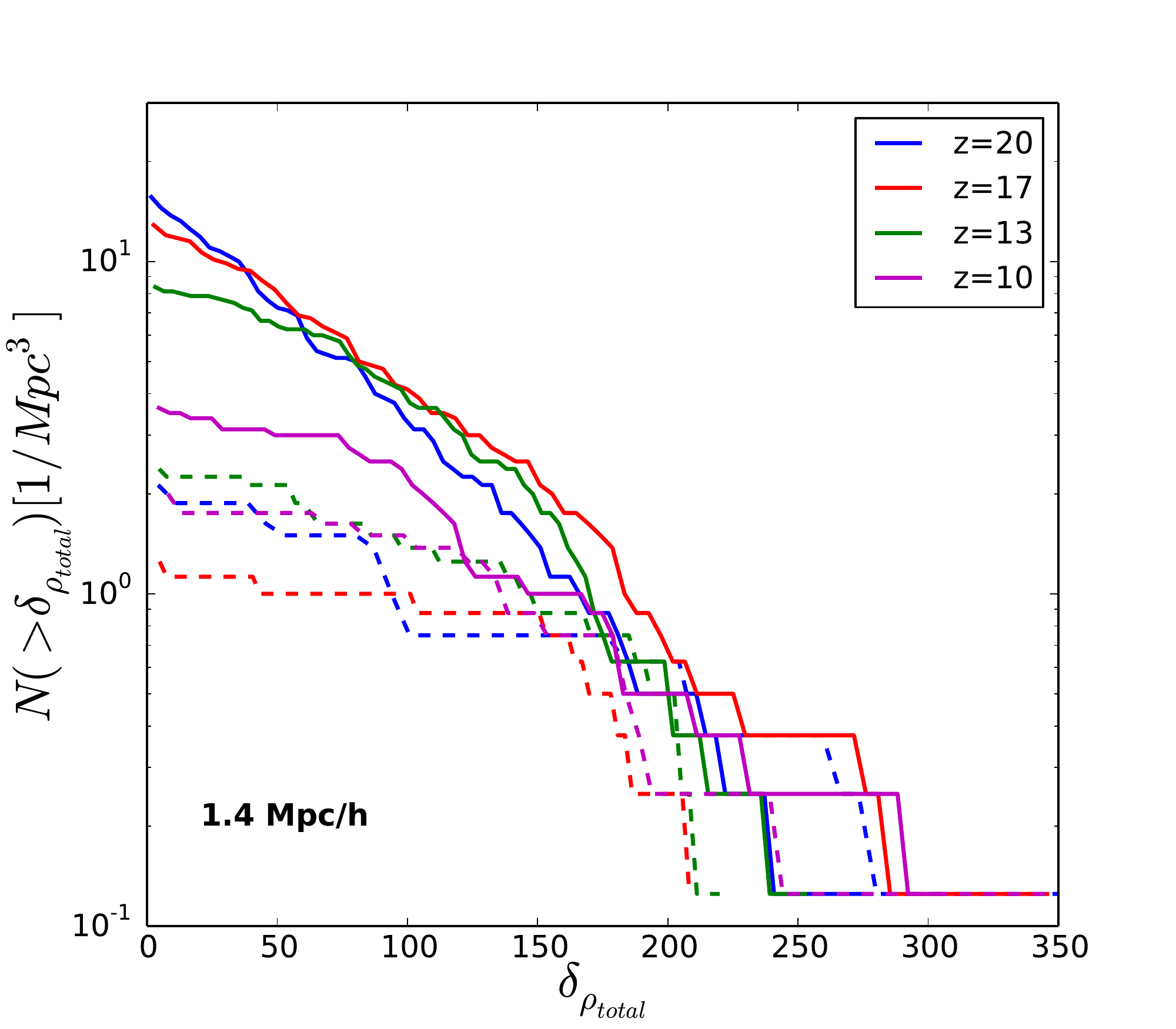}
\end{minipage}\qquad
\begin{minipage}[b]{.4\textwidth}
\centering
\includegraphics[scale=.36]{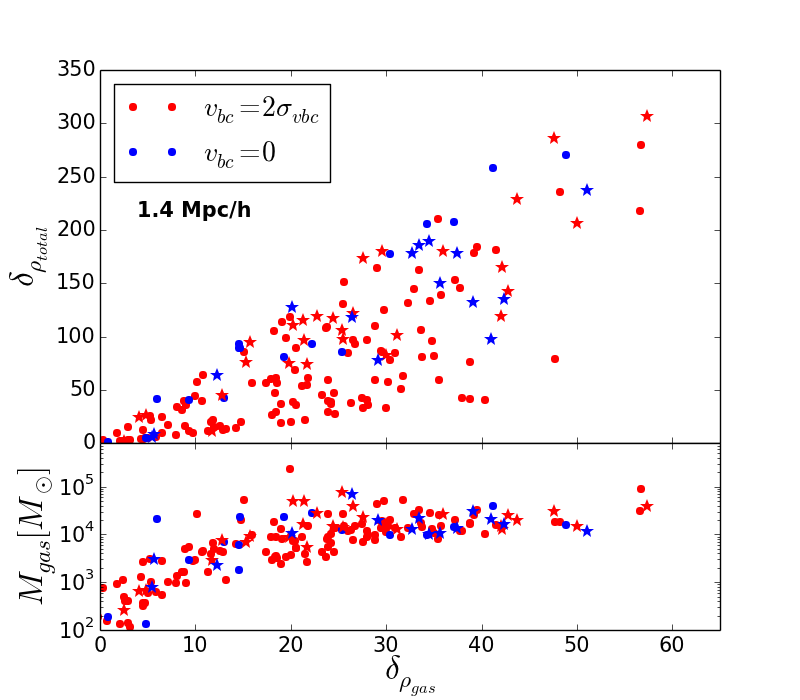}
\end{minipage}
\caption{Left: Number density of Gas-Primary objects identified in the $1.4$ \Mpch\ simulations as a function of over-density $\delta_{\rho_{total}} = (\rho_{total} - \bar{\rho})/\bar{\rho}$. Right: Total matter over-density $\delta_{\rho_{total}}$ as a function of gas over-density $\delta_{\rho_{gas}} = (\rho_{gas} - \bar{\rho})/\bar{\rho}$ (top panel) and gas mass as a function of gas over-density (bottom panel) for Gas-Primary objects identified at redshifts 20 (circles) and 10 (stars). In both plots we only consider Gas-Primary objects located outside the virial radius of the nearest dark matter halo and which have a gas fraction above the cosmic baryonic fraction.}
\label{fig:over-densityGasP}
\end{figure*}

\begin{figure*}
\centering
\begin{minipage}[b]{.4\textwidth}
\centering
\includegraphics[scale=.36]{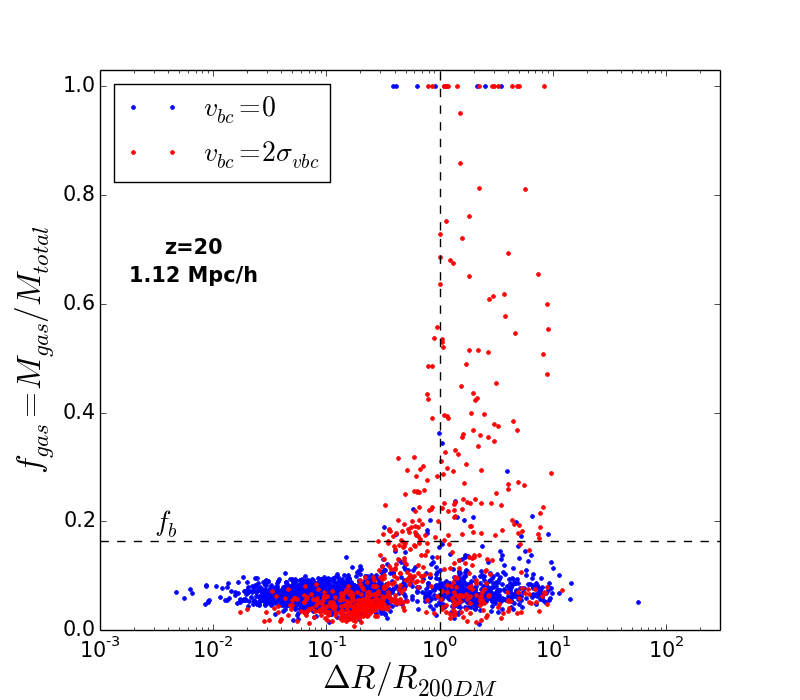}
\end{minipage}\qquad
\begin{minipage}[b]{.4\textwidth}
\centering
\includegraphics[scale=.36]{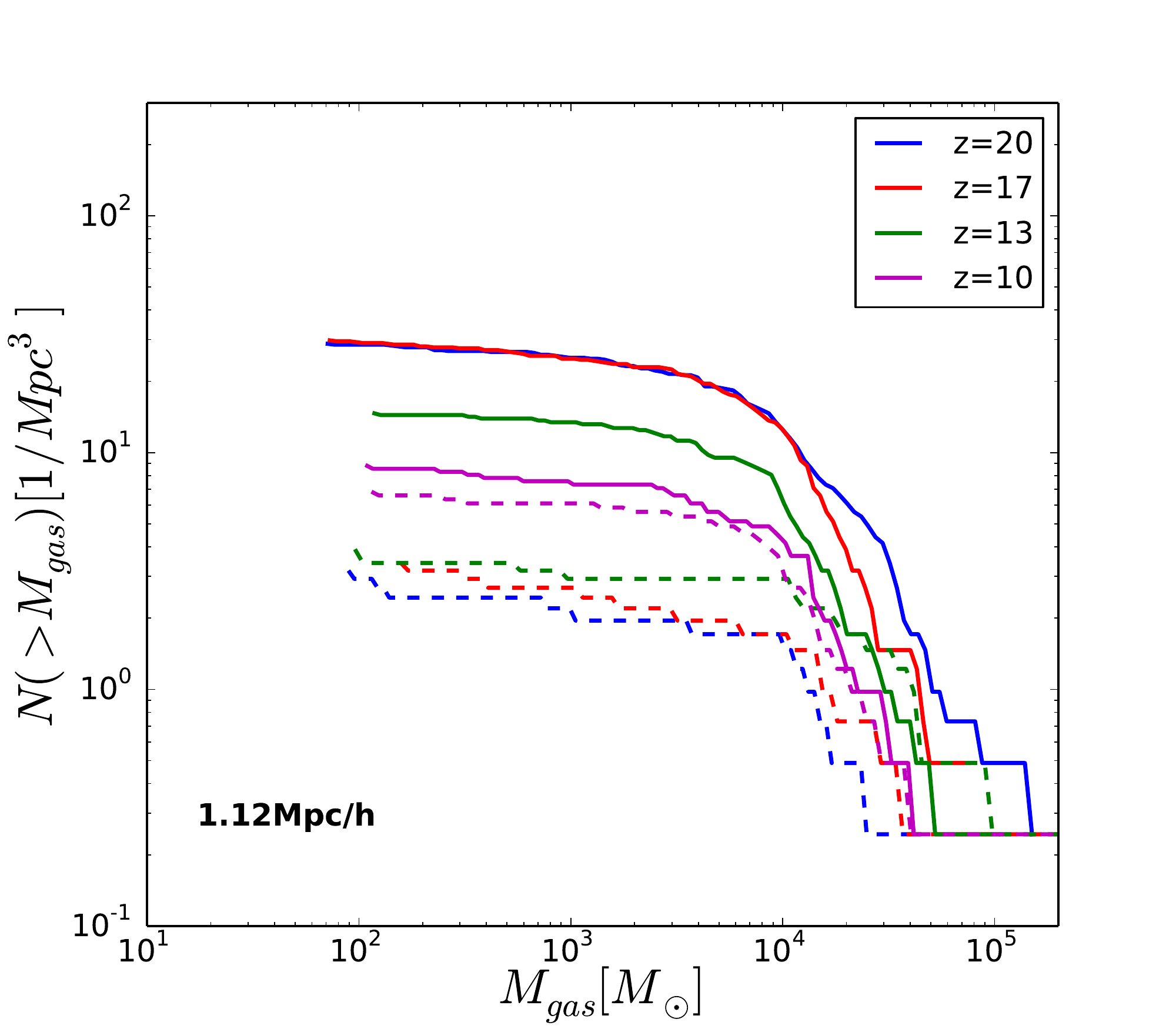}
\end{minipage}
\begin{minipage}[b]{.4\textwidth}
\centering
\includegraphics[scale=.36]{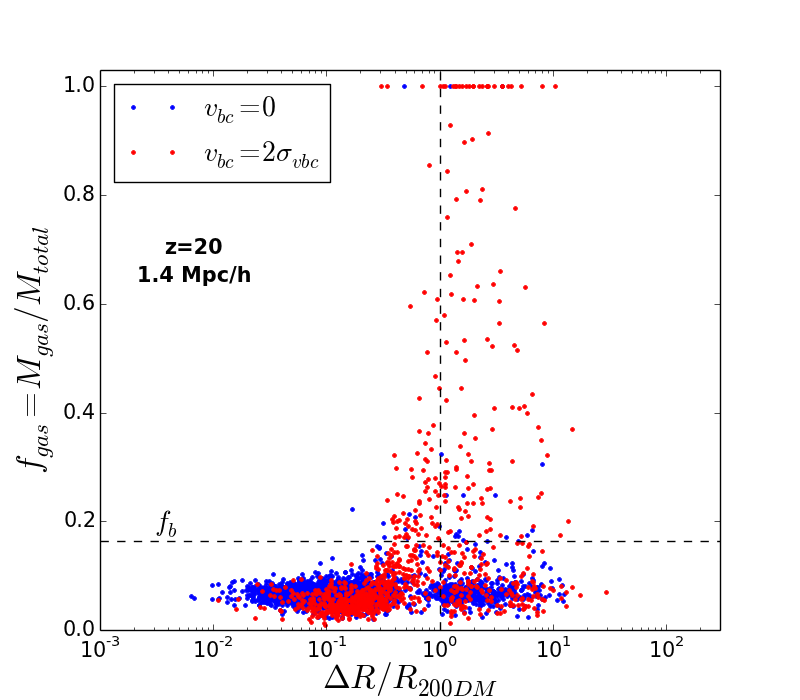}
\end{minipage}\qquad
\begin{minipage}[b]{.4\textwidth}
\centering
\includegraphics[scale=.36]{PaperPlots_reversedLines/HaloMassFunc_Mgas_14Mpc_gfb_gr200.pdf}
\end{minipage}\qquad
\begin{minipage}[b]{.4\textwidth}
\centering
\includegraphics[scale=.36]{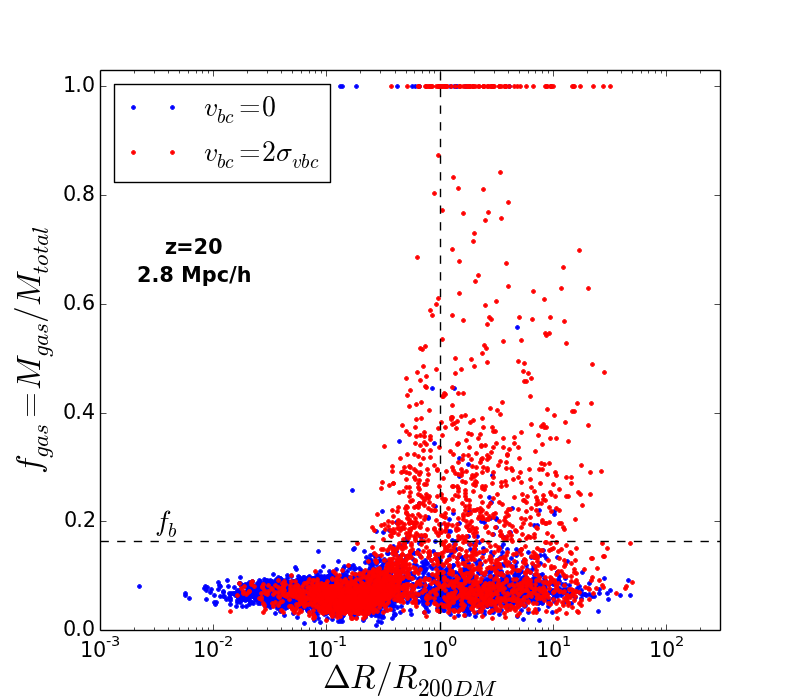}
\end{minipage}\qquad
\begin{minipage}[b]{.36\textwidth}
\centering
\includegraphics[scale=.36]{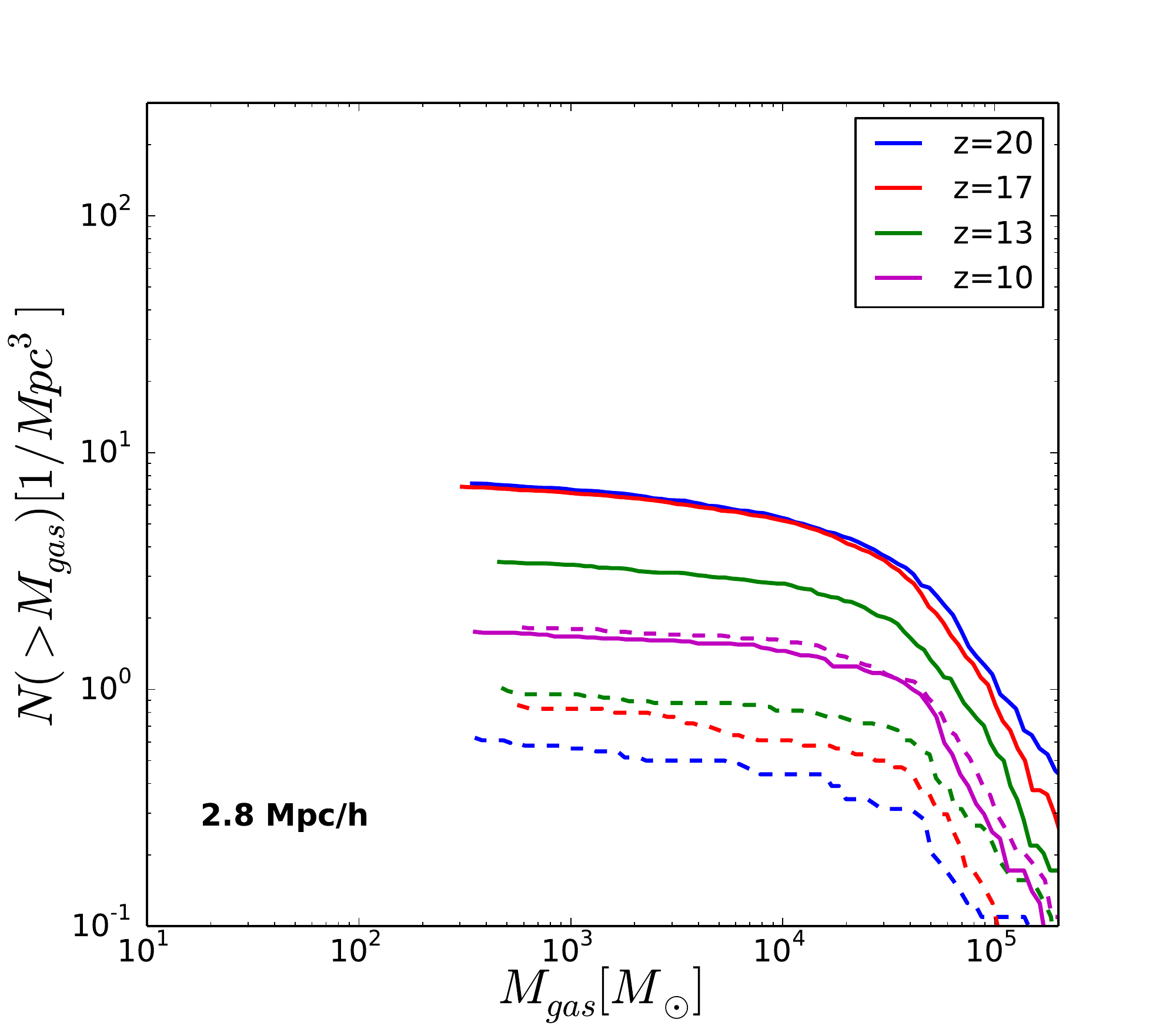}
\end{minipage}\qquad
\caption{Left panels: Gas fraction as a function of the distance to the closest dark matter halo for the Gas-primary objects identified at redshift 20, in the $1.12$ \Mpch\, $1.4$ \Mpch\ and $2.8$ \Mpch\ simulations. Right panels: Halo mass function for the gas abundant gas-primary structures residing outside the virial radius of the closest dark matter halo identified in the lower 3 resolution runs. The solid lines are computed from the high-stream velocity runs, while the dashed lines, from the zero-$v_{bc}$ simulations. All quantities are computed inside of the optimal ellipsoid, as defined in Section \ref{sec:halodef}.}
\label{fig:fgasmgasdeltar1}
\end{figure*}

As mentioned, a good fraction of all the objects identified lie inside the virial radius of the closest dark matter halo and have a gas fraction smaller than the cosmic baryonic fraction. These are the objects we assume to be the gas components of the DM-primary, Gas-secondary objects discussed in Section \ref{sec:DMpGASs}. However, there are also many gas-primary objects located outside the virial radius of the closest dark matter halo. In the case of the zero-stream velocity runs the vast majority of them still have gas fractions lower than the cosmic baryonic fraction, but in the non-zero stream velocity cases we identify a substantial number of objects virtually devoid of dark matter, particularly at high redshift. In the $700$ \kpch\ run with $v_{bc} = 2\times\sigma_{v_{bc}}$, approximately $10\%$($1\%$) of all identified Gas-Primary objects have a gas fraction above $0.8$ at redshift 20(10). As shown in Figure \ref{fig:fgasmgasdeltar1}, the other resolutions explored produce qualitatively similar results, albeit, the number of Gas-Primary objects virtually devoid of dark matter is slightly smaller with only $\sim3\%(0.5\%)$ of them having a gas fraction greater than $0.8$ at redshift $20(10)$. The slight discrepancy in the number of objects with high gas fractions is, in part, due to the fact that in the highest resolution run we can access significantly lighter structures than the other simulations and, when residing outside of dark matter haloes, these low mass objects tend to have a higher gas fraction than their heavier counterparts. 

With decreasing redshift and stream velocity, the number of gas abundant objects located outside the virial radius of dark matter halo decreases. The bottom panels of Figure \ref{fig:fgasmgasdeltar} show the number density distributions of gas-abundant Gas-primary objects residing outside of the virial radius of the closest dark matter halo for a variety of different redshifts. Interestingly, the difference between the abundance of such objects in the zero-stream velocity and non-zero stream velocity simulations decreases with decreasing redshift. It seems that the objects identified at high redshift in the non-zero stream velocity simulations slowly get absorbed inside dark matter haloes, whereas in the zero-stream velocity runs, more such objects appear with time. Qualitatively similar trends are observed in all of our simulations, regardless of the box size used (Figure \ref{fig:fgasmgasdeltar1}, right panels).

A more in-depth analysis of the matter content of the gas abundant objects located outside of dark matter haloes in our simulations, shows relatively high gas column densities (in excess of $10^{21}\, {\rm cm^{-2}}$, Figure \ref{fig:3dEllipsoids_DensityProfiles}), suggesting that under the right circumstances these objects might survive to low redshifts. Furthermore, the average matter density inside of the best fit ellipsoids is generally well in excess of the mean matter density of the universe, indicating that some of these Gas-Primary objects might collapse and evolve outside of dark matter haloes, just as \cite{Naoz:2014bqa} predicted. We find that, while the non-zero stream velocity simulations produce more gas abundant objects located outside of dark matter haloes, they  generally have a smaller over-density than their zero-stream velocity counterparts (Figure \ref{fig:over-densityGasP}, left panel). Furthermore, it seems that it is these objects with lower over-density that end up having a higher gas fraction than the objects identified in the $v_{bc} = 0$ simulations. The right panel in Figure \ref{fig:over-densityGasP} shows the total matter over-density as a function of the gas over-density identified at redshifts 20 (circles) and 10 (stars) in the gas abundant Gas-Primary objects located outside of dark matter haloes. Both the total over-density and the gas over-density are computed with respect to the mean total matter density in the universe. Objects identified in  high stream velocity simulations exhibiting a lower matter over-density enclosed inside tightly fitting ellipsoids tend to have a significantly larger gas over-density than those identified in the zero-stream velocity simulations, while objects with a high matter over-density generally have a similar gas content, independent of the value of the stream velocity. Furthermore, we find that the gas mass contained in these gas abundant objects increases with gas over-density up to what seems to be a large-over-density limit of $\approx 10^5M_{\astrosun}$. Similar results are inferred from all simulations, regardless of resolution. 

We note that, there is a qualitative difference between the gas-primary objects identified in the $v_{bc} = 0$ and the $v_{bc} = 2\times\sigma_{v_{bc}}$ scenarios. All of the gas abundant objects identified in the zero-stream velocity runs, regardless of the resolution of the simulation, are short lived. They generally have been torn out of larger haloes by dynamical processes and fall back into another DM halo shortly thereafter. Therefore, while there may be, at any given time, a handful of gas-abundant objects in the zero-stream velocity simulations, they do not survive for very long. 
However, as suggested by Figures \ref{fig:3dEllipsoids_DensityProfiles}-\ref{fig:over-densityGasP}, these high density system may lead to small scale structure formation (as their over density exceeds unity) and they may harbor dark matter deprived stellar systems.

\begin{table}
\centering
\begin{tabular}{| l | l | r | c | c|}
\hline
$z$ 		& $M_{gas} (M_{\astrosun})$	& $f_{gas}$	& $\frac{\Delta R}{R_{200,DM}}$	& $M_{200,DM} (M_{\astrosun})$  \\\hline 
20	& $6.52\times10^4$	& 38\%	& 1.43	& $1.06\times10^6$ \\
19	& $4.24\times10^4$	& 29\%	& 1.67	& $1.24\times10^6$ \\
18	& $4.71\times10^4$	& 35\%	& 1.37	& $2.04\times10^6$ \\
17	& $9.23\times10^4$	& 36\%	& 1.19	& $2.54\times10^6$ \\
16	& $7.02\times10^3$	& 43\%	& 1.19	& $2.98\times10^6$ \\
15.8	& $1.18\times10^4$	& 49\%	& 1.19	& $3.02\times10^6$ \\
15	& $2.48\times10^4$	& 16\%	& 1.22	& $3.34\times10^6$ \\
14	& $1.25\times10^4$	& 38\%	& 0.87	& $3.53\times10^6$ \\
13	& $7.39\times10^3$	& 27\%	& 0.92	& $4.16\times10^6$ \\
12.6	& $2.55\times10^4$	& 28\%	& 0.86	& $4.19\times10^6$ \\\hline
\end{tabular}
\caption{History of the evolution of one Gas-Primary object in the $700$ \kpch\ 
run with $v_{bc} = 2\times\sigma_{v_{bc}}$. After redshift 12, the object 
evaporates and it cannot conclusively be matched to any other object in the 
simulation. }\label{table:baryClump}
\end{table}

In the $v_{bc} = 2\times\sigma_{v_{bc}}$ runs, the evolution of the gas-primary objects is slightly more complex. While virtually all of the gas-abundant objects found in these simulations at redshift 10 have also been torn out of larger haloes by dynamical processes and we expect them to have a similar fate as the objects found in the zero-$v_{bc}$ simulations, at intermediate and large redshifts we do find a handful of objects which have survived and evolved outside dark haloes for a significant period of time. 
Table \ref{table:baryClump} shows an example of the history of such an object. We first identify the Gas-primary object at redshift 20, when it is living outside of a dark matter halo 30 times its mass. As it evolves towards lower redshifts, it is slowly approaching the dark matter halo and, while the mass of the DM halo is steadily increasing, that of the Gas-primary object exhibits some stochasticity. This is partly due to fact that, being a smaller object, the Gas-primary halo has a harder time accreting or even holding on to the gas it already has. At redshift 15, the companion dark matter halo merges with a smaller dark matter halo and this event sends the Gas-primary object on a spiral towards the resulting halo. By redshift 12, it is completely accreted. 

Nevertheless, objects like the one in Table \ref{table:baryClump} are rare in our high stream velocity simulations and the gas-abundant structures which are located outside of dark matter haloes at high redshift tend to get accreted into a nearby halo on much shorter time scales. The fact that we do identify a few such longer lived objects does indicate that some of them could survive up to the present time, as predicted by \cite{Naoz:2014bqa}, though they do not appear to account for the primary formation mechanism of globular clusters. 

\section{Discussion}\label{sec:discussion}

This study investigates the impact of the streaming motion of baryons with respect to dark matter on the formation and gas content of dark matter haloes. Using high-resolution numerical simulations we study the potential to form long lived, baryon dominated structures in the context of different values of the stream velocity. We account for gravitational forces, hydrodynamic fluxes and adiabatic gas processes to evolve dark matter and baryons from redshift 200 to redshift 10, inside cosmological boxes of up to $2.8$ \Mpch\ on a side. In this way, we systematically quantify the impact of a non-zero stream velocity on the formation and gas content of dark matter haloes spanning more than 4 orders of magnitude in mass. 

Our major results are as follows: 
\begin{enumerate}
\item {\bf Halo abundance.}
We show that a non-zero stream velocity suppresses the formation of dark matter haloes of masses $M <$ few$ \times10^7 M_{\astrosun}$, consistently with previous theoretical and numerical studies. We quantify this suppression for  $M > 10^5M_{\astrosun}$, to be about 60\% at z=20 in our $v_{bc} = 2\times\sigma_{v_{bc}}$ simulations. Furthermore, we show that this suppression persists to lower redshifts and at $z = 10$ it is on the order of 25\% (see Figure \ref{fig:haloMassFnc-DM}).
\item {\bf Gas fraction in dark matter haloes.}
The baryon stream velocity has an impact not only on the halo number density, but also on the gas fraction identified inside dark matter haloes, as shown in \cite{Maio:2010qi, Greif:2011iv, O'Leary:2012rs, Fialkov:2011iw, Naoz:2011if, Naoz:2012fr, Tseliakhovich:2010yw, Dalal:2010yt, Richardson:2013uqa}. Expanding on previous studies we find that the stream velocity suppresses the gas fraction inside haloes with masses up to $\sim 10^8M_{\astrosun}$, whereas haloes with masses above this scale contain a gas fraction insensitive to the value of the stream velocity (see Figures \ref{fig:rhofgasDMHalos}-\ref{fig:agvFgas3ParamBest}).
\item {\bf Gas density in dark matter dominated haloes.}
The baryonic stream velocity also suppresses the gas density recovered in the inner regions of the dark matter haloes (see Figure \ref{fig:agvRhoGasMassBin}). This is, in part, due to the fact that we do not follow radiative processes in our simulations and, hence, the gas density in the inner regions of the halo is determined by the equilibrium of pressure and gravity. Since the non-zero stream velocity acts as an additional pressure term, it is plausible that the gas densities recovered in this case be lower than in the zero-stream velocity regions.
\item {\bf Gas rich structures.}
We identify a population of gas rich structures ($f_{gas}>f_b$) located outside of the virial radius of dark matter haloes (see Figure \ref{fig:fgasmgasdeltar}), as predicted by \cite{Naoz:2014bqa}.  
Unlike dark matter dominated haloes, these gas rich objects have a filamentary structure (see Figures \ref{fig:3dEllipsoids} and \ref{fig:3dEllipsoids_DensityProfiles}).
For the zero stream velocity case, generally these are clumps which have been torn out of dark matter haloes by dynamical processes and which get reabsorbed into dark matter haloes soon thereafter. On the other hand, for the non-zero stream velocity case many of these gas-rich objects are much longer lived and, even though none of the ones we identified in our simulations survive until the final redshift, we do expect that, allowing for star formation, some might. A fraction of these surviving haloes may potentially become the progenitors of globular clusters, as suggested in \cite{Naoz:2014bqa}.
\begin{figure*}
\centering
\begin{minipage}[b]{.38\textwidth}
\centering
\includegraphics[scale=.38]{Plots/fgas_scatter_700kpc_z20.png}
\end{minipage}\qquad
\begin{minipage}[b]{.5\textwidth}
\centering
\includegraphics[scale=.38]{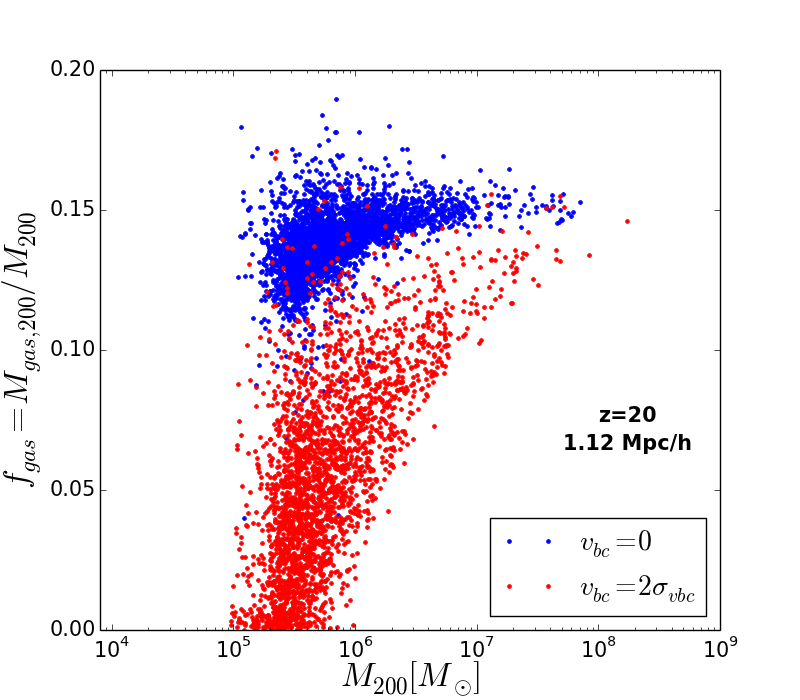}
\end{minipage}\qquad
\begin{minipage}[b]{.4\textwidth}
\centering
\includegraphics[scale=.38]{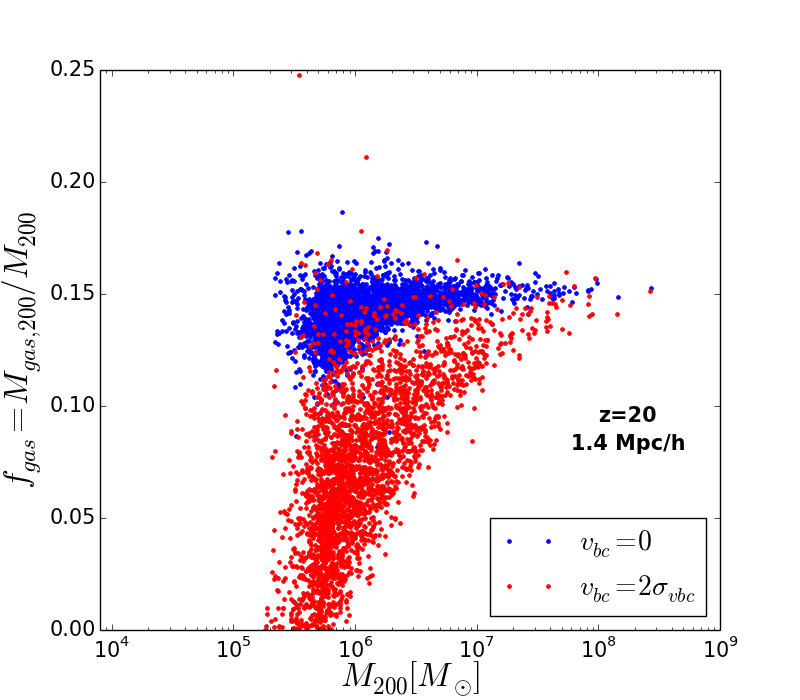}
\end{minipage}\qquad
\begin{minipage}[b]{.43\textwidth}
\centering
\includegraphics[scale=.38]{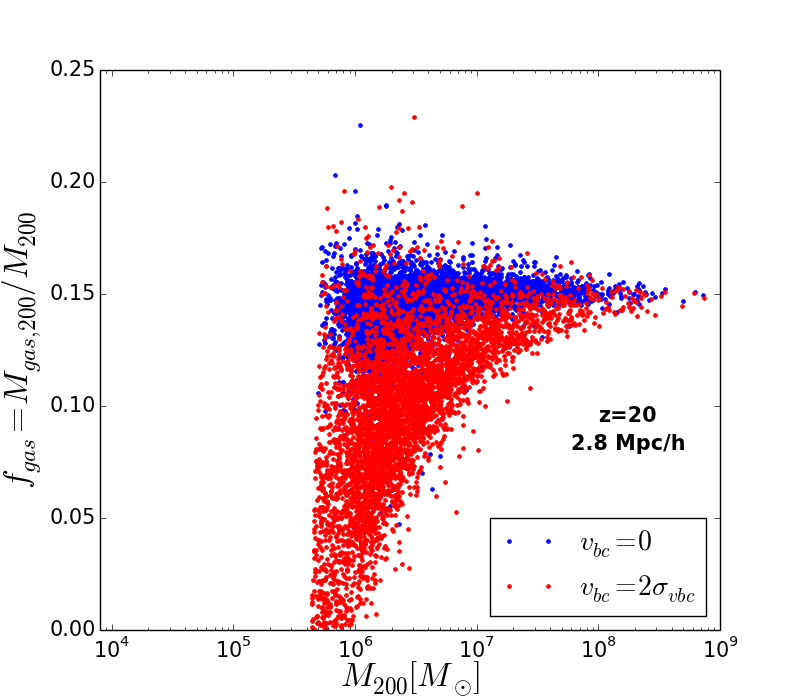}
\end{minipage}
\qquad\caption{Scatter plots of the gas fraction as a function of the total mass of the DM-primary, Gas-secondary objects identified at redshift 20, at the four different resolutions explored by our simulations.}
\label{fig:rhofgasDMHalos1}
\end{figure*}

\item {\bf Gas rich high densities.}
The gas abundant objects we identified residing outside of the virial radius of dark matter haloes do, in fact, reach high densities, suggesting that they are non-linear structures (Figures \ref{fig:3dEllipsoids_DensityProfiles}-\ref{fig:over-densityGasP}). These dense region may have a significant impact on the process of reionization. Moreover, allowing for star formation, these high-density structures may be the sites of early star formation becoming dark matter deprived stellar systems, as hypothesized by \cite{Naoz:2014bqa}.
\end{enumerate}


\section*{Acknowledgements}
We thank Ramesh Narayan for useful discussions and Volker Springel for giving us access to {\sc arepo}. SN acknowledges partial support from a Sloan Foundation Fellowship. MV acknowledges support through an MIT RSC award. The simulations were performed on the joint MIT-Harvard computing cluster supported by MKI and FAS.

\bibliographystyle{mnras}
\bibliography{paper_draft5FS}

\label{lastpage}

\end{document}